\documentclass[lettersize,journal]{IEEEtran}
\usepackage{amsmath,amsfonts}
\usepackage{algorithm}
\usepackage{array}
\usepackage[caption=false,font=normalsize,labelfont=sf,textfont=sf]{subfig}
\usepackage{textcomp}
\usepackage{stfloats}
\usepackage{url}
\usepackage{verbatim}
\usepackage{graphicx}
\usepackage{cite}
\usepackage{amssymb}
\usepackage{amsthm}
\usepackage{cleveref}
\usepackage{booktabs}
\usepackage{algpseudocode}
\usepackage{multirow}
\usepackage{subfig}

\newtheorem{theorem}{Theorem}

\newtheorem{corollary}{Corollary}
\newtheorem{definition}{Definition}

\begin{document}

\title{LeakyCLIP: Extracting Training Data from CLIP}

\author{
    Yunhao Chen$^{\dagger}$,
    Shujie Wang$^{\dagger}$,
    Xin Wang,
    Ran He,
    Xingjun Ma $^{*}$,
    and Yu-Gang Jiang $^{*}$%
    \thanks{
        $\dagger$ Equal contribution, $^{*}$ Corresponding author.
    }
    \thanks{
        Yunhao Chen, Shujie Wang and Xin Wang are with the College of Computer Science and Artificial Intelligence, Xingjun Ma is with the Institute of Trustworthy Embodied AI, Yu-Gang Jiang is with the College of Intelligent Robotics and Advanced Manufacturing, 
        Fudan University, Shanghai, China, Ran He is with the Institute of Automation, Chinese Academy of Sciences
        (e-mail: 24110240013@m.fudan.edu; 24110240084@m.fudan.edu.cn; xinwang22@m.fudan.edu.cn; rhe@nlpr.ia.ac.cn; xingjunma@fudan.edu.cn; ygj@fudan.edu.cn).
    }
    
}


\maketitle

\begin{abstract}
Understanding the memorization and privacy leakage risks in Contrastive Language--Image Pretraining (CLIP) is critical for ensuring the security of multimodal models.  Recent studies have demonstrated the feasibility of extracting sensitive training examples from diffusion models, with conditional diffusion models exhibiting a stronger tendency to memorize and leak information. In this work, we investigate data memorization and extraction risks in CLIP through the lens of \emph{CLIP inversion}, a process that aims to reconstruct training images from text prompts. To this end, we introduce \emph{LeakyCLIP}, a novel attack framework designed to achieve high-quality, semantically accurate image reconstruction from CLIP embeddings.
We identify three key challenges in CLIP inversion: 1) non-robust features, 2) limited visual semantics in text embeddings, and 3) lack of low-level features in reconstructions. To address these challenges, LeakyCLIP employs 1) adversarial fine-tuning to enhance optimization smoothness, 2) linear transformation-based embedding alignment, and 3) controlled Stable Diffusion-based refinement to improve fidelity. 
Empirical results demonstrate the superiority of LeakyCLIP, achieving over 258\% improvement in Structural Similarity Index Measure (SSIM) for ViT-B-16 compared to baseline methods on LAION-2B subset. Furthermore, we uncover a pervasive leakage risk, showing that training data membership can even be successfully inferred from the metrics of low-fidelity reconstructions. Our work introduces a practical method for CLIP inversion while offering novel insights into the nature and scope of privacy risks in multimodal models.

\end{abstract}


\section{Introduction}
\label{sec:intro}

\IEEEPARstart{W}{ith} the widespread adoption of large models across diverse domains \cite{kaddour2023challenges,meng2024application,chen2024comprehensive,zhou2024twindiffusion,ma2025safety}, concerns about their ability to memorize and leak raw training data have grown significantly \cite{wei2024memorization,pittaluga2023ldp}. Recent research has successfully demonstrated that training data can be extracted from large language models (LLMs) \cite{carlini2021extracting,carlini2023extracting} and text-to-image diffusion models \cite{somepalli2023diffusion,somepalli2023understanding,gu2023memorization}. These studies further reveal that conditional training—such as incorporating class labels or text prompts—amplifies memorization risks in diffusion models \cite{somepalli2023understanding,gu2023memorization}. 

Motivated by these findings, we extend the investigation to Contrastive Language-Image Pretraining (CLIP) \cite{radford2021learning}, a class of models designed to align images and detailed text descriptions through contrastive learning. By design, CLIP conditions each image on its corresponding textual description (and vice versa), establishing a bidirectional dependency that could potentially intensify memorization risks. Given CLIP’s widespread adoption in downstream tasks, we investigate whether this cross-modal conditioning similarly facilitates—or potentially amplifies—training data extraction. Beyond its practical implications, such an investigation also deepens our understanding of memorization in CLIP models, shedding light on their privacy risks.

To investigate this, we focus on a specific type of data extraction attack known as a \textbf{model inversion} attack \cite{fredrikson2014privacy}. A model inversion attack aims to reconstruct the original input data— in this case, images from the training set—given access to the model and potentially some auxiliary information \cite{nguyen2023re}. When applied to CLIP, this attack translates into a \textbf{CLIP inversion} task.
In this paper, we formalize CLIP inversion as a task to reconstruct an image conditioned on its paired text prompt from the training data. In this paper, we propose LeakyCLIP, a novel data extraction attack for CLIP models that enables reconstructed images to achieve high fidelity.

\begin{figure*}
    \centering
    \includegraphics[width=0.7\linewidth]{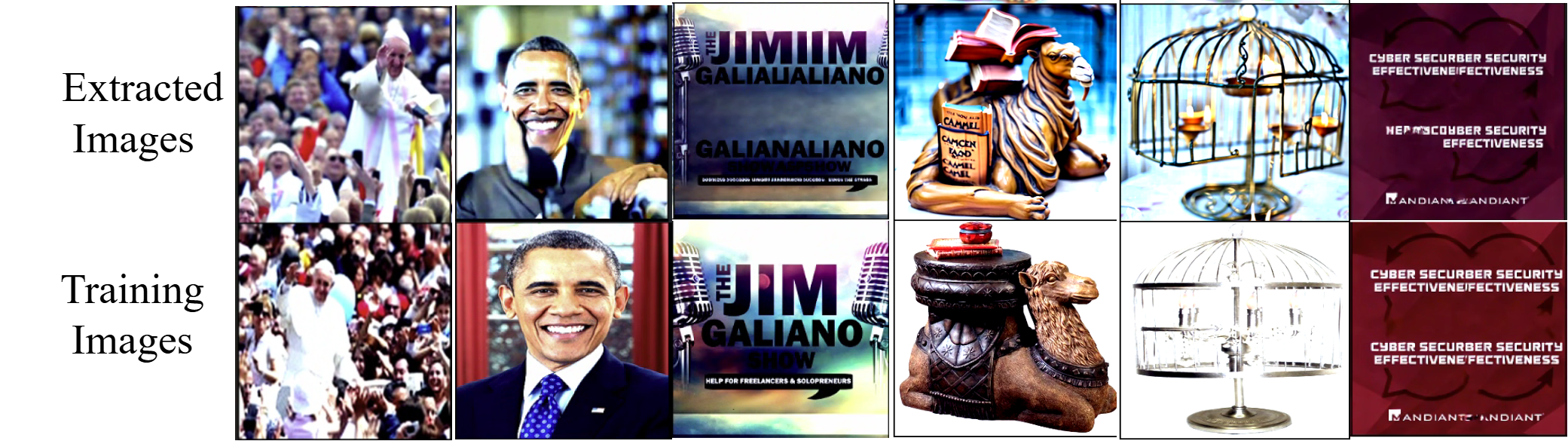}
    
    \caption{The first row presents reconstructed images generated by LeakyCLIP, while the second row shows the corresponding training images. According to our Highly Similar metric, these reconstructions are recognized as highly similar to the originals. }
    \label{fig:demo}
    
\end{figure*}

Our work on CLIP inversion \cite{kazemi2024we} identifies and addresses a series of challenges in order. First, we observe that the primary problem lies in the \textbf{non-robust optimization landscape} during model inversion process (see Figure \ref{fig:optim_smooth}). CLIP learns features that are highly predictive but may not correspond to meaningful visual concepts \cite{ilyas2019adversarial,ganz2023perceptually}. This leads to an unstable optimization process that distorts gradients and makes it hard to achieve good perceptual alignment. To solve this problem, we first employ \textbf{adversarial fine-tuning}, which stabilizes the inversion process and creates a robust feature space for building the image.

Once the optimization process was stable, a new issue emerges: although the robust text features are reliable, they exhibit \textbf{limited visual semantics}. While effective at capturing abstract concepts, they lack the high-level visual information required to produce coherent images, such as object layout and scale.  To address this, we introduce an \textbf{embedding space alignment} technique by learning a linear transformation between CLIP’s image and text embeddings, projecting the robust text features into pseudo-image embeddings that better approximate true image representations.

Though this pseudo-image embedding provides good visual semantics, it still lacks the fine-grained, low-level features needed for a realistic image. This limitation arises because the pseudo-image embedding itself is constructed from robust text embeddings, which inherently lacks low-level visual features. To address this issue, we propose a \textbf{diffusion-based refinement} step with a pre-trained Stable Diffusion model \cite{rombach2022high}. Each refinement step refines the reconstructed images by perturbing them with Gaussian noise and applying a reverse diffusion process using \textbf{Stable Diffusion} \cite{rombach2022high}. To prevent the diffusion model from leaking its own memorized content, we carefully control the process. At each denoising step, we ensure the refined image remains highly similar to the image generated in the alignment stage. 
This forces the diffusion model to add only low-level details—such as textures and sharp edges—without leaking its contents into the reconstructed images.
Extensive experiments validate its effectiveness, with example results shown in \cref{fig:demo,fig:visual_examples} .

Using LeakyCLIP, we demonstrate that privacy leakage extends well beyond high-fidelity reconstructions. Even for low-fidelity ones, the reconstruction metrics can still reliably infer whether an image was included in the training set, indicating a far more pervasive privacy threat. These risks become more consequential in fine-tuning and downstream deployment scenarios, where CLIP models are often adapted using private or proprietary image–text pairs (e.g., enterprise data or user uploads).

 In summary, our contributions are threefold:
\begin{itemize}
     \item We introduce \textbf{LeakyCLIP}, a novel model inversion attack method for CLIP models, which significantly improves the quality of reconstructed images compared to existing approaches. Specifically, for the ViT-B-16 model on a subset of LAION-2B, our method achieves a 258\% improvement in SSIM compared with the baseline method.
    \item We identify three key challenges in performing inversion attacks on CLIP models and propose targeted solutions to address each. Through empirical evaluation, we demonstrate the effectiveness of our methods.
    \item  We demonstrate a pervasive privacy risk beyond high-fidelity reconstructions, showing that the evaluation metrics from even poor reconstructions can be used to infer training data membership successfully.
\end{itemize}

\section{Related Work}

\paragraph{CLIP}\cite{radford2021learning} is a type of VLM used for various tasks, including classification, captioning, retrieval and generation. It consists of vision and text modules that map images and text to corresponding embeddings. Using contrastive learning, CLIP minimizes the distance between embeddings of matched image-text pairs and maximizes it for mismatched ones. Since its introduction, CLIP has been widely adopted in applications such as text-guided image synthesis \cite{nichol2021glide}, high-resolution image generation in diffusion models \cite{rombach2022high, zhou2024twindiffusion}, and image segmentation \cite{luddecke2022image}. However, its wide use has raised concerns about privacy and security, including issues like identity inference attacks \cite{hintersdorf2024does}, backdoor attacks \cite{bai2024badclip}, bias, NSFW content \cite{kazemi2024we}, and training data memorization \cite{jayaraman2024d}. This work explores the privacy risks of CLIP, with a particular focus on memorization and data extraction.

\paragraph{Memorization and Model Inversion Risks in CLIP}
The widespread adoption of Vision-Language Models (VLMs) and their reliance on high-quality data ~\cite{xu2023demystifying} have raised concerns about their privacy, particularly regarding the memorization of training data. Recent work has begun to quantify this phenomenon formally. Recent works~\cite{wang2025captured,wang2024memorization,webster2025multi,xiu2025caprecover,wei2025open} reveal that CLIP tends to memorize text and image data, especially the face identities. One of the most direct ways to demonstrate the tangible risk of such memorization is through \textbf{Model Inversion (MI) attacks}, which aim to reconstruct the private training data. MI attacks have a rich history, evolving from early methods targeting classifiers with low-dimensional data~\cite{fredrikson2014privacy, fredrikson2015model} to more advanced, GAN-based techniques capable of recovering high-dimensional images from both white-box~\cite{zhang2020secret} and black-box~\cite{kahla2022label} models. Despite these advancements~\cite{yuan2022secretgen, yuan2023pseudo, struppek2022plug}, most methods are tailored for classifiers and are less effective against complex, multimodal models like CLIP. While recent work by Kazemi et al.~\cite{kazemi2024we} demonstrated initial success in extracting data from CLIP, a deep exploration of the technical challenges remained absent. To fill this gap, we investigate the specific challenges in CLIP inversion, including noisy gradients from non-robust features~\cite{ilyas2019adversarial, ganz2023perceptually}, the semantic gap between modalities and the lack of low-level features in reconstructions. Based on these, we propose \textbf{LeakyCLIP}, a novel framework that combines adversarial fine-tuning, embedding alignment, and generative refinement to form a new extraction attack.

\section{Proposed Method}

\begin{figure*}[!ht]
    
    \centering
    \includegraphics[width=1\linewidth]{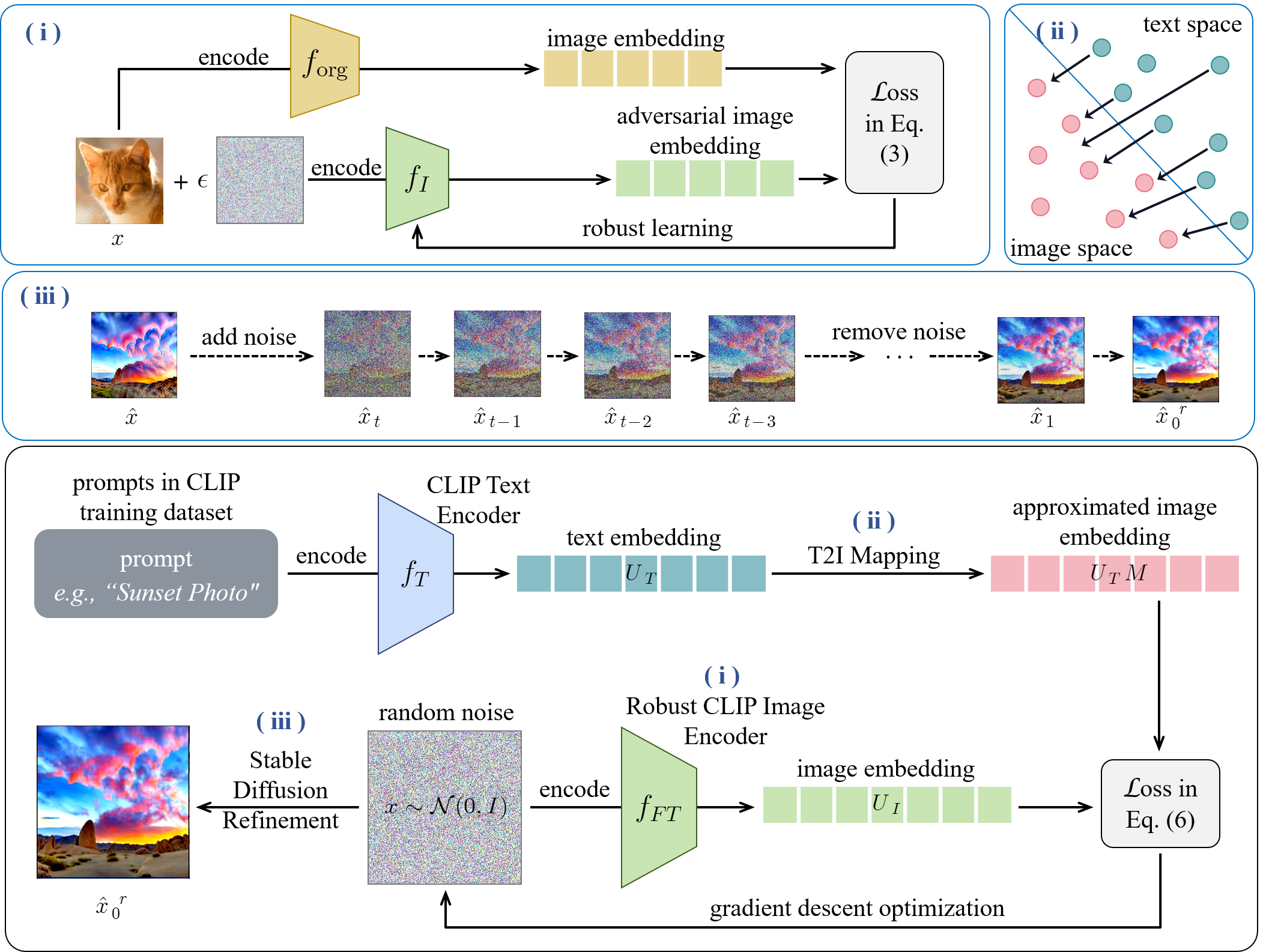}
    
    \caption{LeakyCLIP for Training Data Extraction. (i) Adversarial Fine-Tuning (AFT):  The image encoder $f_I$ is Adversarial Fine-Tuned to smooth the optimization landscape. $f_{org}$ is a duplicated image encoder whose parameters are frozen during fine-tuning. (ii) Embedding Alignment (EA): A linear transformation matrix $M$ is learned to align the text and image embeddings, mapping text embedding $U_T$ to an approximate image embedding $U_T M$. (iii) Diffusion Refinement (DR):The reconstructed image is refined using Stable Diffusion to improve visual quality.}
    \label{fig:framework}
    
\end{figure*}

\subsection{Threat model}
The attacker is assumed to have white-box access to the CLIP parameters. 
The considered threat model is well-established and essential for rigorously probing the privacy vulnerabilities of large-scale models. Assuming white-box access to model parameters is a standard practice in model inversion attacks \cite{fredrikson2015model,kazemi2024we,fang2024privacy}. These assumptions are justified by the inherent difficulty of inverting complex models designed to generalize\cite{ma2025safety,fang2024privacy}, making this framework essential for understanding the upper bounds of potential privacy leakage from sophisticated adversaries. Even under this white-box assumption, the baseline produces reconstructions with very poor quality, indicating the task remains highly non-trivial. Our contributions focus on addressing the intrinsic difficulty of large-scale pretrained vision-language model inversion in this setting, rather than relaxing access assumptions.

\subsection{CLIP Inversion}

CLIP inversion refers to the task of reconstructing a training
image from its textual description. Given an image–text pair
$(x, t)$ from the training set, the goal is to reconstruct an
image $\hat{x}$ that closely matches $x$, using the text $t$ and
the CLIP model. The objective of the current CLIP inversion method \cite{kazemi2024we} is formulated as follows:
\begin{align}
\label{baseline_img}
\hat{x} = \underset{x}{\arg\min} \left(1- \frac{ f_I(x)^\top \mathbf{u}_T}{\| f_I({x}) \|_2 \| \mathbf{u}_T \|_2} + \lambda \mathcal{L}_{\text{TV}}(x) \right),
\end{align}
where \( f_I \) is the CLIP image encoder, \( \mathbf{u}_T \) is the text embedding for \( t \), and \( \mathcal{L}_{\text{TV}}(x) \) is the Total Variation (TV)\cite{kazemi2024we} loss with hyperparameter \( \lambda \) controlling its strength. The first term in \cref{baseline_img} maximizes the cosine similarity between the reconstructed image embedding and the text embedding, while the second is for fidelity.





Although the current CLIP inversion methods can generate semantically relevant images, it often fails to produce high-quality reconstructions. We identify three key factors contributing to this issue. First, the image encoder's lack of robustness introduces noise into the gradients, resulting in an unsmooth optimization landscape. Second, unlike image embeddings, text embeddings primarily capture semantic content and lack the high-dimensional and fine-grained visual information required for accurate image reconstruction. Finally, the reconstructed images are not sufficiently refined, leading to a loss of fine details and reduced fidelity.

To further understand CLIP inversion and its connection to training data memorization, \textbf{we introduce in the \cref{sec:thm} a theoretical metric that quantifies CLIP’s memorization capacity for a specific dataset. Using this framework, we prove that more detailed textual descriptions amplifies the risk of training data leakage.} Our theoretical findings are supported by experimental validation.

\subsection{Adversarial Fine-Tuning for Smoothed Optimization}

Based on \cref{baseline_img}, there is a clear correlation between the smoothness of the loss function and that of the image encoder $f_I$. A regular (non-robust) encoder with insufficient smoothness can lead to an unstable optimization process. To improve the encoder's smoothness, we apply an unsupervised adversarial tuning method, Fine-tuning for Adversarially Robust Embeddings (FARE)\cite{schlarmann2024robustclip}, to the image encoder \( f_I \) in \cref{eq:linear_img}. The adversarial fine-tuning loss is defined as:
\begin{equation}
f_{FT} = \underset{f_I}{\mathrm{argmin}} \sum_{i=1}^{n} \mathcal{L}_{FARE}(f_I, x_i),
\end{equation}
where the adversarial loss for each data point is given by:
\begin{equation}
\mathcal{L}_{FARE}(f_I, x) = \underset{||z-x||_{\infty} \leq \epsilon}{\max} ||f_I(z) - f_{\text{org}}(x)||_2^2.
\end{equation}
Here, $f_{\text{org}}$ denotes a frozen copy of the original CLIP image encoder before AFT, and $x$ denotes an image from the auxiliary dataset (ImageNet, with any CLIP pretraining data removed). This loss function enforces the perturbed feature vectors \( f_I(z) \) to be closed to the original encoder features \( f_{\text{org}}(x) \), leading to a smoother image encoder. The AFT does not leak more data on the test distribution because its fine-tuning auxiliary dataset is disjoint from the training data. AFT improves results by improving the optimization landscape. We validate the smoothness of the gradients in the experimental section. In addition, AFT also helps mitigate the adverse effects caused by noise commonly present in image datasets. The following theorem formalizes this effect.

\begin{theorem}
\label{thm:AFT}
Let $f_I:\mathbb{R}^d\rightarrow\mathbb{S}^{k-1}$ denote the image encoder. 
Assume that $f_I$ is bi-Lipschitz, i.e., there exist constants $0<m\le M<\infty$ such that
\[m \|x-y\| \le \|f_I(x)-f_I(y)\|
\le M\|x-y\|,
\forall x,y\in\mathbb{R}^d.
\]
Given sample $x\sim p(x)$, we perturb the input with noise $\xi$
\[
\tilde{x}=x+\xi, \ \mathbb{E}[{\| \xi \|}^2] \leq d \sigma^2
\]
and define the reconstruction (consistent with \cref{baseline_img} on $\mathbb{S}^{k-1}$)
\[ \hat{x} = \arg\min_{z}
\|f_I(z)-f_I(\tilde{x})\|^2. \]
Let $\hat{p}$ denote the distribution induced by $\hat{x}$, it satisfies
\[ W_2^2(p,\hat{p}) \le
\frac{4d\sigma^2 M^2}{m^2}, \]
where $W_2$ denotes the 2-Wasserstein distance.
\end{theorem}
Specifically, as shown in \cref{fig:lipschitz_ratio}, AFT primarily decreases the upper Lipschitz constant $M$ of the encoder, making the encoder robust to noise, resulting in a better optimization landscape. While the lower Lipschitz constant $m$, which characterizes feature distinguishability, changes much less in practice. Consequently, $\frac{M}{m}$ becomes smaller after AFT, leading to a tighter upper bound on the Wasserstein distance between the original and reconstructed data distributions.

\subsection{Linear Transformation Based Embedding Alignment}

With a more stable optimization process, the next challenge is the limited visual information in text embeddings. While robust, these text features are not sufficient on their own to create a visually coherent image. To address the gap between text and image embeddings, we follow the framework in \cite{tan2024contrastive} and revisit the graph-based view of CLIP. This view implies a \emph{coupled linear relation} between text and image embeddings, as outlined in \cref{th:text2image}. Motivated by this relation, we learn a lightweight, data-driven linear mapping on an auxiliary dataset to approximate image embeddings for reconstruction.

\begin{theorem}
\label{th:text2image}
Let \( A = \{a_1, \dots, a_n\} \) and \( B = \{b_1, \dots, b_m\} \) represent sets of text and image nodes, respectively. The weight of the adjacency matrix \( \mathbf{W} \) is computed using cosine similarity. The text node degree matrices \( \mathbf{D}_T \) and the image node degree matrices \( \mathbf{D}_I \) are diagonal. The matrix \( \mathbf{U}_I \) represents image embeddings while \( \mathbf{U}_T \) represents text embeddings. Then the CLIP text and image embeddings satisfy the following linear relation:
\begin{align}
\label{eq:linear_relation}
\mathbf{U}_I (\mathbf{I} - \Lambda)
= \mathbf{D}_I^{-1/2} \mathbf{W}^\top \mathbf{D}_T^{-1/2} \mathbf{U}_T,
\end{align}
where \( \Lambda \) contains the $d$ most important eigenvalues corresponding to the selected dimensions.
\end{theorem}

The proof can be found in the \cref{sec:thm}. The relation in \cref{eq:linear_relation} establishes a global linear dependence between the stacked text and image embeddings through the graph operators \( \mathbf{D}_T, \mathbf{D}_I, \mathbf{W} \) and the spectrum \( \Lambda \). Because this dependence involves all training samples jointly, we could not compute it in practice. We therefore approximate the induced operator using a global linear surrogate matrix \( M \), learned by minimizing the following reconstruction error:
\begin{equation}
\label{M_cal}
M = \underset{M \in \mathbb{R}^{d \times d}}{\arg\min}
\left\| \mathbf{U}_I^{\text{aux}} - \mathbf{U}_T^{\text{aux}} M \right\|_F^2,
\end{equation}
where \( \mathbf{U}_I^{\text{aux}} \) and \( \mathbf{U}_T^{\text{aux}} \) are auxiliary matrices of image and text embeddings, and \( \| \cdot \|_F \) denotes the Frobenius norm. The optimal surrogate matrix \( M \) is obtained via the Moore--Penrose pseudo-inverse as
\[
M = (\mathbf{U}_T^{\text{aux}})^\dagger \mathbf{U}_I^{\text{aux}}.
\]
Using this learned surrogate, the image embeddings are \emph{approximately} predicted from text embeddings as
\[
\hat{\mathbf{u}}_I = \mathbf{u}_T M.
\]

This approximate linear mapping improves semantic alignment between text and image embeddings in practice, leading to higher-quality reconstructions. The optimization objective is then updated as:
\begin{align}
\label{eq:linear_img}
\hat{x} = \underset{x}{\arg\min}
\left(
1 -
\frac{ f_{FT}(x)^\top \hat{\mathbf{u}}_I}{\| f_{FT}(x) \|_2 \, \| \hat{\mathbf{u}}_I \|_2}
+ \lambda \mathcal{L}_{\text{TV}}(x)
\right) .
\end{align}

\subsection{Stable Diffusion Based Refinement}

Finally, even with good visual semantics, the reconstruction still lack the low-level features of a realistic image. This is because the pseudo-image embeddings are themselves derived from text embeddings, which inherently lack low-level visual information. To address this, we introduce Stable Diffusion~\cite{rombach2022high} via SDEdit~\cite{meng2022sde} purely as an \emph{image-space denoiser} for refinement.

Starting with the reconstructed image $\hat{x}$ from Eq.~(6), where the vision encoder is replaced with $f_{\mathrm{FT}}$, we first add Gaussian noise to obtain a noisy image $\hat{x}_T$ at step $T$. The reverse process of the pre-trained diffusion model $p_\theta$ is then applied to denoise the image:
\begin{equation}
  p_\theta(\hat{x}_{t-1} \mid \hat{x}_t)
  =
  \mathcal{N}\bigl(
    \hat{x}_{t-1};\,
    \mu_\theta(\hat{x}_t, t),
    \sigma^2(t)\mathbf{I}
  \bigr),
  \label{eq:sd_reverse_uncond}
\end{equation}
where $\hat{x}_t$ is the noisy image at step $t$, and $\mu_\theta(\hat{x}_t, t)$ is the predicted mean for denoising. The iterative process reduces noise while refining the image’s visual details.

In score-based form, the denoising can be written as
\begin{equation}
  \hat{x}_{t-1}
  =
  \hat{x}_t
  -
  \Delta_t \,
  \nabla_{\hat{x}_t}
  \log p_\theta(\hat{x}_t),
  \label{eq:sd_score_uncond}
\end{equation}
where $\nabla_{\hat{x}_t} \log p_\theta(\hat{x}_t)$ is the gradient of the log-likelihood of $\hat{x}_t$ under the fixed, pre-trained diffusion prior. In our implementation, we do not provide any text prompt or CLIP embedding to the diffusion model; the only input is the reconstructed image $\hat{x}$ and added Gaussian noise.

To prevent the diffusion model from introducing its own memorized content, we implement a strict control mechanism. At each step $t$ of the iterative denoising process, after calculating the potential refined image $\hat{x}_{t-1}$, we perform a similarity check against the original input reconstruction $\hat{x}$:
\begin{equation}
  \text{Sim}(\hat{x}_{t-1}, \hat{x}) \ge \tau.
  \label{eq:sd_similarity_constraint}
\end{equation}
Here, $\text{Sim}(\cdot, \cdot)$ is a similarity function and $\tau$ is a pre-defined threshold. We set $\tau=1$ in our implementation. Writing the function this way is simply to keep the step generic so other similarity functions and thresholds can be plugged in, whereas our implementation uses the HS metric. If this condition is not met, we terminate the refinement procedure and use the image from the previous step, $\hat{x}_t$, as the final output. This ensures the process does not drift from the input provided by the inversion stage. We use the Highly Similar Metric as the similarity function which is detailed in Section ~\ref{sec:experiments}.

Thus, CLIP inversion (AFT+EA) serves as the primary extraction mechanism, while this controlled diffusion refinement acts solely as a post-processing step that improves local visual fidelity. Qualitative examples in the  \cref{fig:visual_examples}  further show that this constrained process (DR) mainly sharpens textures and reduces artifacts, while successfully preserving the global semantics produced by AFT+EA.

\subsection{Complete Procedure} 

LeakyCLIP (\cref{fig:framework}) operates in three stages: 
\textbf{(1) Adversarial Fine-tuning}: Adversarially fine-tune CLIP’s image encoder \( f_{org} \) via FARE \cite{schlarmann2024robustclip} to smooth gradients;  
\textbf{(2) Linear Transformation Based Embedding Alignment}: Map text embeddings \( \mathbf{u}_T \) to pseudo-image embeddings \( \mathbf{u}_T M \) using a learned linear matrix \( M \) (Theorem~\ref{th:text2image});  
\textbf{(3) Stable Diffusion Based Refinement}: Reconstruct images using gradient descent on \( f_{FT} \), then refine with Stable Diffusion \cite{rombach2022high} to improve fidelity.  
Overall, this pipeline combines gradient-based inversion with generative refinement to extract high-fidelity training data from CLIP models. In addition, we provide a detailed pseudo-code of the  algorithm in the \cref{alg:leakyclip}.

\newcommand{\AlgNote}[1]{%
    \Statex \hspace{\algorithmicindent}{\footnotesize\emph{// #1}}%
}

\begin{algorithm}[t]
\caption{LeakyCLIP Training Data Extraction}
\label{alg:leakyclip}
\small
\begin{algorithmic}[1]

\Require Text prompt $t$
\Require CLIP encoders $(f_{\mathrm{org}}, f_I, f_T)$
\Require Auxiliary set $\mathcal{D}_{\mathrm{aux}}=\{(x_i,t_i)\}_{i=1}^{n}$
\Require Diffusion model $p_\theta$
\Require Hyperparameters $\epsilon,\lambda,\eta,K,T,\tau$
\Ensure Reconstructed image $\hat{x}_0$

\Statex \textbf{Stage 1: Adversarial fine-tuning}

\For{each minibatch $\{x_i\}_{i=1}^{B}\subset\mathcal{D}_{\mathrm{aux}}$}

    \AlgNote{Frozen target embeddings}
    \State $
    u_i^{\mathrm{org}} \gets f_{\mathrm{org}}(x_i)
    $

    \AlgNote{Adversarial perturbation}
    \State $
    \delta_i^\star
    =
    \arg\max_{\|\delta\|_\infty\leq\epsilon}
    \left\|
    f_I(x_i+\delta)-u_i^{\mathrm{org}}
    \right\|_2^2
    $

    \State $
    z_i \gets x_i+\delta_i^\star
    $

    \AlgNote{AFT objective}
    \State $
    \mathcal{L}_{\mathrm{AFT}}
    =
    \sum_{i=1}^{B}
    \left\|
    f_I(z_i)-u_i^{\mathrm{org}}
    \right\|_2^2
    $

    \State $
    f_I
    \gets
    f_I-\eta\nabla_{f_I}\mathcal{L}_{\mathrm{AFT}}
    $

\EndFor

\State $
f_{\mathrm{FT}}\gets f_I
$

\Statex \textbf{Stage 2: Embedding alignment}

\AlgNote{Auxiliary embedding matrices}
\State $
U_T
\gets
\left[
f_T(t_1),\ldots,f_T(t_n)
\right]^\top
$

\State $
U_I
\gets
\left[
f_{\mathrm{FT}}(x_1),\ldots,f_{\mathrm{FT}}(x_n)
\right]^\top
$

\AlgNote{Least-squares text-to-image mapping}
\State $
M
=
\arg\min_{M}
\left\|
U_I-U_TM
\right\|_F^2
$

\State $
M\gets U_T^\dagger U_I
$

\State $
u_T \gets f_T(t)
$

\State $
\hat{u}_I \gets u_T M
$

\Statex \textbf{Stage 3: CLIP inversion}

\State $
x^{(0)}\sim\mathcal{N}(0,I)
$

\For{$k=1,\ldots,K$}

    \AlgNote{Embedding alignment loss}
    \State $
    \mathcal{L}_{\mathrm{cos}}(x)
    =
    1-
    \cos
    \left(
    f_{\mathrm{FT}}(x),
    \hat{u}_I
    \right)
    $

    \AlgNote{Inversion objective}
    \State $
    \mathcal{L}_{\mathrm{inv}}(x)
    =
    \mathcal{L}_{\mathrm{cos}}(x)
    +
    \lambda\mathcal{L}_{\mathrm{TV}}(x)
    $

    \State $
    x^{(k)}
    =
    x^{(k-1)}
    -
    \eta
    \left.
    \nabla_x
    \mathcal{L}_{\mathrm{inv}}(x)
    \right|_{x=x^{(k-1)}}
    $

\EndFor

\State $
\hat{x}\gets x^{(K)}
$

\Statex \textbf{Stage 4: Controlled diffusion refinement}

\AlgNote{Forward noising}
\State $
\hat{x}_T
=
\sqrt{\bar{\alpha}_T}\hat{x}
+
\sqrt{1-\bar{\alpha}_T}\epsilon_T,
\quad
\epsilon_T\sim\mathcal{N}(0,I)
$

\For{$s=T,\ldots,1$}

    \AlgNote{Reverse denoising}
    \State $
    \epsilon_\theta
    \gets
    \epsilon_\theta(\hat{x}_s,s),
    \quad
    z_s\sim\mathcal{N}(0,I)
    $

    \State $
    \tilde{x}_{s-1}
    =
    \frac{1}{\sqrt{\alpha_s}}
    \left(
    \hat{x}_s
    -
    \frac{\beta_s}{\sqrt{1-\bar{\alpha}_s}}
    \epsilon_\theta
    \right)
    +
    \sigma_s z_s
    $

    \AlgNote{Similarity constraint}
    \If{$\mathrm{Sim}(\tilde{x}_{s-1},\hat{x}) < \tau$}
        \State \textbf{break}
    \Else
        \State $
        \hat{x}_{s-1}\gets \tilde{x}_{s-1}
        $
    \EndIf

\EndFor

\State $
\hat{x}_0 \gets \hat{x}_{s}
$

\State \Return $\hat{x}_0$

\end{algorithmic}
\end{algorithm}

\section{Experiments}
\label{sec:experiments}
\subsection{Experimental Settings}

\noindent\textbf{Datasets}\;
We evaluate our method on three datasets: a subset of LAION-2B (5000 samples) \cite{kirstain2024laionhd}, the Furniture Object Dataset \cite{abrarlohia_sample_furniture_object}, and a subset of Flickr30k (5000 samples) \cite{jia2015guiding}. ImageNet \cite{5206848} is used for adversarial fine-tuning of CLIP models via FARE. The LAION-2B dataset provides a diverse, real-world benchmark for evaluating scalability and performance under challenging conditions. In contrast, the Furniture Object Dataset offers specialized content, enabling assessment of domain-specific feature extraction capabilities. Additionally, Flickr30k, with its human-annotated captions, facilitates evaluation of model performance on abstract textual prompts. Together, these datasets ensure comprehensive testing across varying scales, content domains, and text complexity levels. 
Over 98\% of the images in the datasets mentioned above are included in the training data of the CLIP model we used.

\begin{figure*}[!ht]
    
    \centering
    \includegraphics[width=0.8\linewidth]{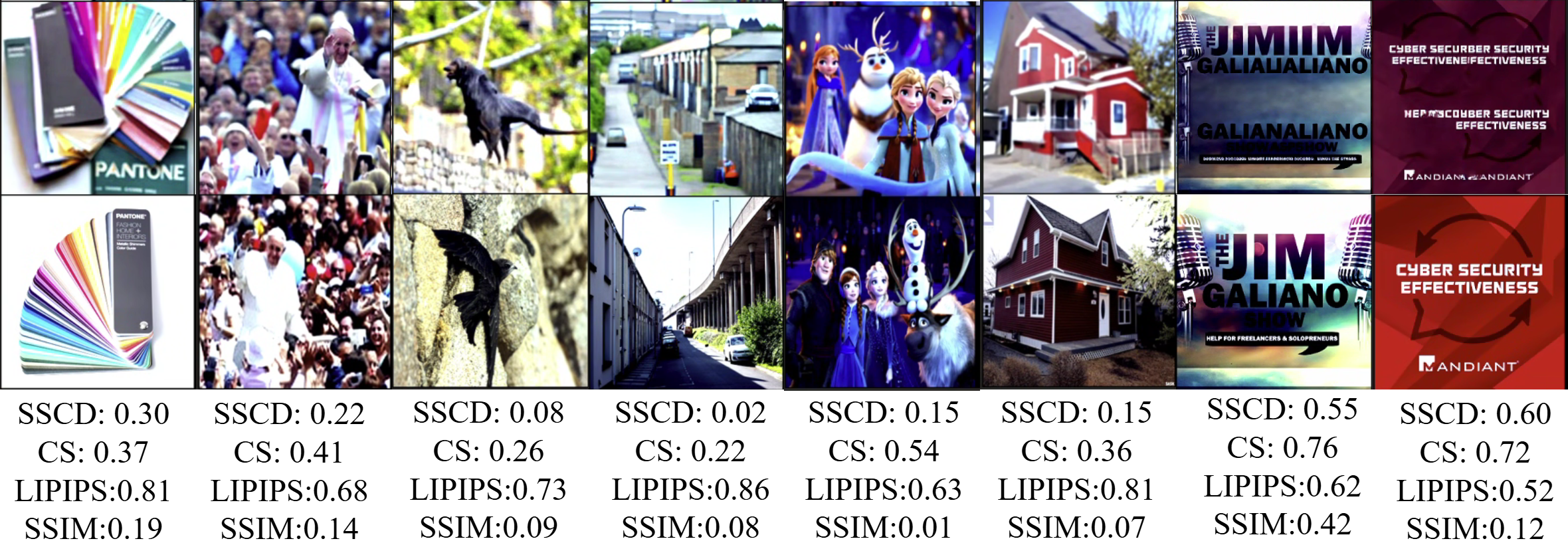}
    
    \caption{Top: Reconstructed images by LeakyCLIP. Bottom: The original images and metric values.}
    \label{fig:metric_show}
\end{figure*}

\noindent\textbf{Models}\;
We apply LeakyCLIP to widely used CLIP image encoders, including ViT-L-14 , ViT-B-16, and ViT-B-32 from OpenCLIP implementation \cite{ilharco_gabriel_2021_5143773}. 

\noindent\textbf{Baseline Method} Our baseline is adopted from \cite{kazemi2024we}, which optimizes \cref{baseline_img}. To the best of our knowledge, this is the only existing method for CLIP inversion capable of reconstructing semantically meaningful images. 

\noindent\textbf{Implementation Details}
For AFT, we follow the hyperparameters and settings from RobustVLM \cite{schlarmann2024robustclip}. Adversarial examples are generated using a 10-step PGD attack \cite{madry2017towards} with an $L_{\infty}$ norm bound of $\epsilon = 4/255$ (in short eps=4) and used to fine-tune the model on ImageNet. For text-to-image mapping, we randomly select 2,000 text-image pairs from LAION-2B dataset (disjoint with inversion dataset) to compute the mapping matrix $M$ for each model. 
For model inversion, we use AdamW as the optimizer, with a learning rate of 0.175, gradient clipping at 0.001, and 200 epochs per reconstruction. We use Stable Diffusion 2 for diffusion-based refinement of reconstructed images, with an image strength of 0.55, 50 denoising steps. These parameters align with established defaults in prior work \cite{von-platen-etal-2022-diffusers,schlarmann2024robustclip,kazemi2024we}.

\subsection{Evaluation Metrics}
\label{subsec:eval_metrics}

\begin{table*}[ht]
\caption{Ablation study results showing the performance impact of individual LeakyCLIP components. HS is reported in percentage points (e.g., 4.20 denotes 4.20\%).}
\centering
\small
\resizebox{\textwidth}{!}{%
\begin{tabular}{l l c c c c c c c c c c c c c c c}
    \toprule
     \multirow{2}{*}{\textbf{Model}} & \multirow{2}{*}{\textbf{Method}} & \multicolumn{5}{c}{\textbf{LAION-2B Subset}} & \multicolumn{5}{c}{\textbf{Furniture Object Dataset}} & \multicolumn{5}{c}{\textbf{Flickr30k Subset}}  \\
    \cmidrule(r){3-7} \cmidrule(r){8-12} \cmidrule(r){13-17}
    & & SSIM \(\uparrow\) & LPIPS \(\downarrow\) & CS \(\uparrow\) & SSCD \(\uparrow\) &HS(\%) \(\uparrow\) & SSIM \(\uparrow\) & LPIPS \(\downarrow\) & CS \(\uparrow\) & SSCD \(\uparrow\) &HS(\%) \(\uparrow\) & SSIM \(\uparrow\) & LPIPS \(\downarrow\) & CS \(\uparrow\) & SSCD \(\uparrow\) &HS(\%)\(\uparrow\)\\
    \midrule
    \multirow{8}{*}{\textbf{ViT-B-16}}
    & Baseline  & 0.042 & 0.973 & 0.406 & 0.010 & 0.06 & 0.048 & 0.991 & 0.369 & 0.019 & 0.00 & 0.029 & 0.977 & 0.445 & 0.011 & 0.00 \\
    & EA  & 0.046 & 0.956 & 0.282 & 0.019 & 0.00 & 0.054 & 0.965 & 0.280 & 0.025 & 0.00 & 0.031 & 0.963 & 0.302 & 0.011 & 0.00\\
    & DR  & 0.059 & 0.939 & 0.337 & 0.013 & 0.86 & 0.073 & 0.952 & 0.272 & 0.013 & 0.00 & 0.039 & 0.918 & 0.379 & 0.023 & 0.04 \\
    & DR+EA  & 0.061 & 0.908 & 0.350 & 0.022 & 0.64 & 0.076 & 0.921 & 0.361 & 0.022 & 0.60 & 0.040 & 0.884 & 0.427 & 0.033 & 0.08 \\
    & AFT & 0.093 & 0.914 & 0.406 & 0.036 & 0.00 & 0.118 & 0.904 & 0.410 & 0.043 & 0.00 & 0.052 & 0.924 & 0.421 & 0.031 & 0.00 \\
    & AFT+EA  & 0.121 & 0.861 & 0.328 & 0.034 & 1.04 & 0.160 & 0.852 & 0.384 & 0.032 & 0.60 & 0.058 & 0.883 & 0.332 & 0.025 & 0.16 \\
    & AFT+DR  & 0.112 & 0.859 & 0.395 & 0.049 & 0.18 & 0.146 & 0.852 & 0.372 & 0.052 & 0.00 & 0.058 & 0.866 & 0.434 & 0.057 & 0.02 \\
    & AFT+DR+EA & \textbf{0.151} & \textbf{0.819} & \textbf{0.462} & \textbf{0.065} & \textbf{4.20} & \textbf{0.199} & \textbf{0.820} & \textbf{0.486} & \textbf{0.055} & \textbf{3.61} & \textbf{0.065} & \textbf{0.850} & \textbf{0.502} & \textbf{0.060} & \textbf{1.42} \\ 
    \midrule
    \multirow{8}{*}{\textbf{ViT-B-32}}
    & Baseline  & 0.046 & 0.979 & 0.411 & 0.018 & 0.00 & 0.053 & 0.992 & 0.370 & 0.019 & 0.00 & 0.030 & 0.988 & 0.451 & 0.015 & 0.00 \\
    & EA  & 0.048 & 0.976 & 0.278 & 0.017 & 0.00 & 0.060 & 0.989 & 0.282 & 0.025 & 0.00 & 0.031 & 0.982 & 0.286 & 0.017 & 0.00\\
    & DR  & 0.064 & 0.927 & 0.362 & 0.019 & 0.02 & 0.079 & 0.931 & 0.316 & 0.018 & 0.00 & 0.039 & 0.912 & 0.420 & 0.034 & 0.04 \\
    & DR+EA  & 0.063 & 0.921 & 0.379 & 0.025 & 0.38 & 0.087 & 0.924 & 0.394 & 0.021 & 0.00 & 0.039 & 0.895 & 0.440 & 0.033 & 0.00 \\
    & AFT & 0.098 & 0.920 & 0.397 & 0.038 & 0.00 & 0.128 & 0.908 & 0.395 & 0.048 & 0.00 & 0.052 & 0.937 & 0.422 & 0.037 & 0.00 \\
    & AFT+EA  & 0.122 & 0.884 & 0.319 & 0.042 & 0.12 & 0.163 & 0.883 & 0.368 & 0.035 & 0.00 & 0.055 & 0.905 & 0.314 & 0.024 & 0.02 \\
    & AFT+DR & 0.113 & 0.853 & 0.390 & 0.052 & 0.44 & 0.154 & 0.842 & 0.378 & 0.058 & 0.00 & 0.058 & 0.864 & 0.418 & 0.064 & 0.28 \\
    & AFT+DR+EA & \textbf{0.144} & \textbf{0.809} & \textbf{0.487} & \textbf{0.067} & \textbf{3.46} & \textbf{0.201} & \textbf{0.822} & \textbf{0.484} & \textbf{0.062}  & \textbf{3.01} & \textbf{0.061} & \textbf{0.834} & \textbf{0.509} & \textbf{0.071} & \textbf{1.78} \\ 
    \midrule
    \multirow{8}{*}{\textbf{ViT-L-14}}
    & Baseline & 0.041 & 0.970 & 0.408 & 0.014 & 0.00 & 0.047 & 0.982 & 0.364 & 0.007 & 0.00 & 0.028 & 0.974 & 0.434 & 0.004 & 0.00 \\
    & EA & 0.042 & 0.967 & 0.233 & 0.016 & 0.00 & 0.050 & 0.973 & 0.201 & 0.014 & 0.00 & 0.028 & 0.983 & 0.234 & -0.003 & 0.00 \\
    & DR  & 0.057 & 0.951 & 0.276 & 0.009 & 0.00 & 0.070 & 0.961 & 0.231 & -0.006 & 0.00 & 0.037 & 0.938 & 0.351 & 0.015 & 0.00 \\
    & DR+EA  & 0.056 & 0.935 & 0.288 & 0.011 & 0.00 & 0.067 & 0.946 & 0.274 & 0.007 & 0.00 & 0.034 & 0.934 & 0.305 & 0.016 & 0.00 \\
    & AFT & 0.075 & 0.924 & 0.408 & 0.028 & 0.00 & 0.092 & 0.920 & 0.398 & 0.032 & 0.00 & 0.044 & 0.933 & 0.424 & 0.011 & 0.00 \\
    & AFT+EA & 0.082 & 0.911 & 0.271 & 0.027 & 0.00 & 0.100 & 0.902 & 0.284 & 0.043 & 0.00 & 0.043 & 0.928 & 0.266 & 0.009 & 0.00 \\
    & AFT+DR & 0.090 & 0.897 & 0.374 & 0.028 & 0.22 & 0.123 & 0.889 & 0.348 & 0.030 & 0.00 & 0.052 & 0.890 & 0.424 & 0.033 & 0.02 \\
   & AFT+DR+EA & \textbf{0.125} & \textbf{0.803} & \textbf{0.417} & \textbf{0.037} & \textbf{0.84} & \textbf{0.121} & \textbf{0.840} & \textbf{0.408} & \textbf{0.042} & \textbf{0.60} & \textbf{0.047} & \textbf{0.826} & \textbf{0.440} & \textbf{0.052} & \textbf{0.24} \\
    \midrule 
    \multirow{1}{*}{SD} 
    & Sampling & 0.104 & 0.811 & 0.338 & 0.025 & 0.46 & 0.130 & 0.850 & 0.262 & 0.012 & 0.00 & 0.046 & 0.837 & 0.352 & 0.021 & 0.02 \\
    \bottomrule
    \end{tabular}
}
    
    \label{tab:methods_comparison_all}

\end{table*}

\begin{figure*}[ht]
    
    \centering
    \includegraphics[width=1.1\linewidth]{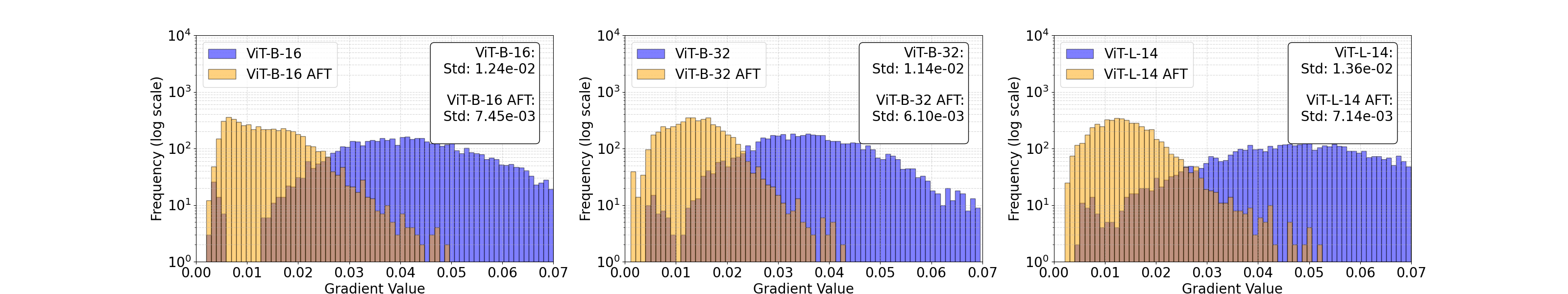}
    
    \caption{Gradient Distribution Comparison: Histograms showing that adversarially fine-tuned CLIP have smaller and less variable gradient norms, reflecting a smoother optimization landscape.}
    \label{fig:optim_smooth}
\end{figure*}

The goal of CLIP inversion is to reconstruct training images from textual descriptions. Unlike traditional model inversion attacks targeting class labels, CLIP inversion operates on text embeddings. We adapt following metrics(examples are shown in \cref{fig:demo} and \cref{fig:metric_show}.):
\begin{itemize}
      
    \item \textbf{SSIM} \cite{wang2004image} $\uparrow$: measures structural similarity between original and reconstructed images. Range: $[-1, 1]$.
    
    \item \textbf{LPIPS} \cite{zhang2018unreasonable} $\downarrow$: assesses perceptual similarity in deep features. Range: $[0, \infty)$.
    
    \item \textbf{CS (CLIP Score)} $\uparrow$: computes cosine similarity between reconstructed and original image embeddings. Range: $[-1, 1]$.
    
    \item \textbf{SSCD} \cite{pizzi2022self} $\uparrow$: measures self-supervised copy detection score using a neural network trained to identify near-duplicate images regardless of transformations such as cropping, resizing, or compression. Range: $[0, 1]$.
    
    \item  \textbf{HS (Highly Similar)} $\uparrow$: a composite measure using thresholds (SSCD $\ge$ 0.4, LPIPS $\le$ 0.55, SSIM $\ge$ 0.6, CS $\ge$ 0.6) as indicators of reconstruction fidelity. The thresholds were determined via human evaluation: five annotators assessed 1000 reconstructions, and thresholds were selected based on majority-voted positives. HS rates of $\sim$4--5\% indicate a \emph{non-trivial minority} of high-fidelity reconstructions. Crucially, given that CLIP was trained on billions of images (e.g., LAION-2B contains 2.3 billion samples).
    
\end{itemize}




\subsection{Experimental Results}

\label{sec:exp_results}
\noindent\textbf{Main Results}\;
We evaluate LeakyCLIP on three datasets: LAION-2B Subset, Furniture Object Dataset, and Flickr30k Subset. Results are summarized in \cref{tab:methods_comparison_all}. On the LAION-2B Subset with ViT-B-16, LeakyCLIP achieves substantial gains: SSIM increases by 258\%, LPIPS decreases by 16\%, CLIP Score rises by 15\%, SSCD improves by 446\%, and HS increases from 0\% to 4.2\%. Similar improvements are observed for ViT-B-32 and ViT-L-14 across all metrics. 

\noindent\textbf{Results Across Datasets}\;
LeakyCLIP achieves the most pronounced improvements on the Furniture Object Dataset, benefiting from its structured content. In contrast, improvements on Flickr30k are more modest, likely due to its abstract and narrative captions.

\noindent\textbf{Results Across CLIP Variants}\;
As shown in \cref{tab:methods_comparison_all}, performance for ViT-L-14 lags behind ViT-B-16 and ViT-B-32, attributable to its larger parameter count (~304M vs. ~88M), which complicates the optimization of the model inversion.

\begin{figure}[h]
    
    \centering
    \includegraphics[width=\columnwidth]{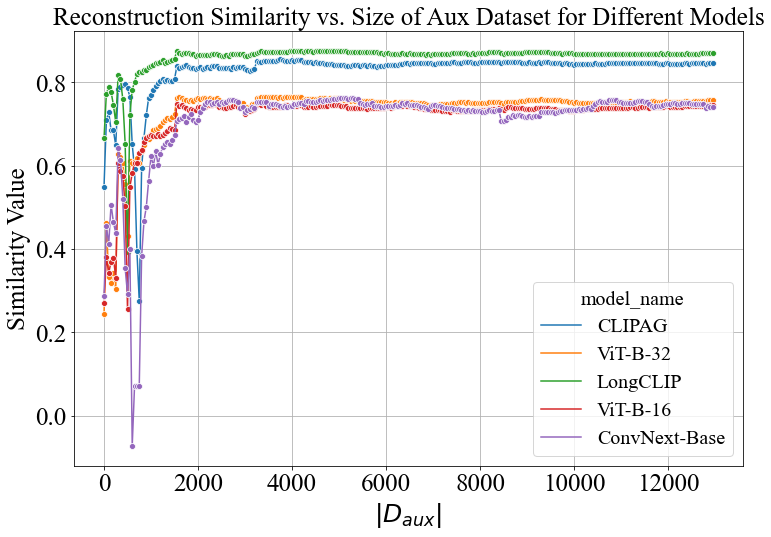}
    
    \caption{The relationship between the reconstruction similarity of image embedding in $D_{test}$ and size of the $D_{aux}$.}
    \label{fig:selection}
\end{figure}

\noindent\textbf{Ablation Study}\;
Ablation analysis of LeakyCLIP's key components---Embedding Alignment (EA), Adversarial Fine-Tuning (AFT), and Diffusion Refinement (DR)---shows that AFT is essential for effective inversion, while DR further enhances perceptual quality (SSIM, LPIPS). The combination of EA with AFT and DR yields the greatest overall boost. Notably, DR or EA alone, or DR+EA without AFT, provide limited improvements, underscoring AFT's central role. Regarding the interesting trend in CS scores: the baseline explicitly optimizes an embedding-matching objective in the attacked CLIP space (cosine similarity), and can therefore overfit to shortcut directions that are highly predictive in CLIP features but weakly grounded in true image semantics. Each component alone disrupts part of this shortcut fitting by stabilizing optimization (AFT), encouraging visual semantics (EA), or improving pixel realism (DR), causing CS to drop before semantics are fully recovered. Only when these techniques are combined can the framework reconstruct images that are semantically faithful, thereby improving CS across evaluation backbones.

\begin{figure}
    
    \centering
    \includegraphics[width=\columnwidth]{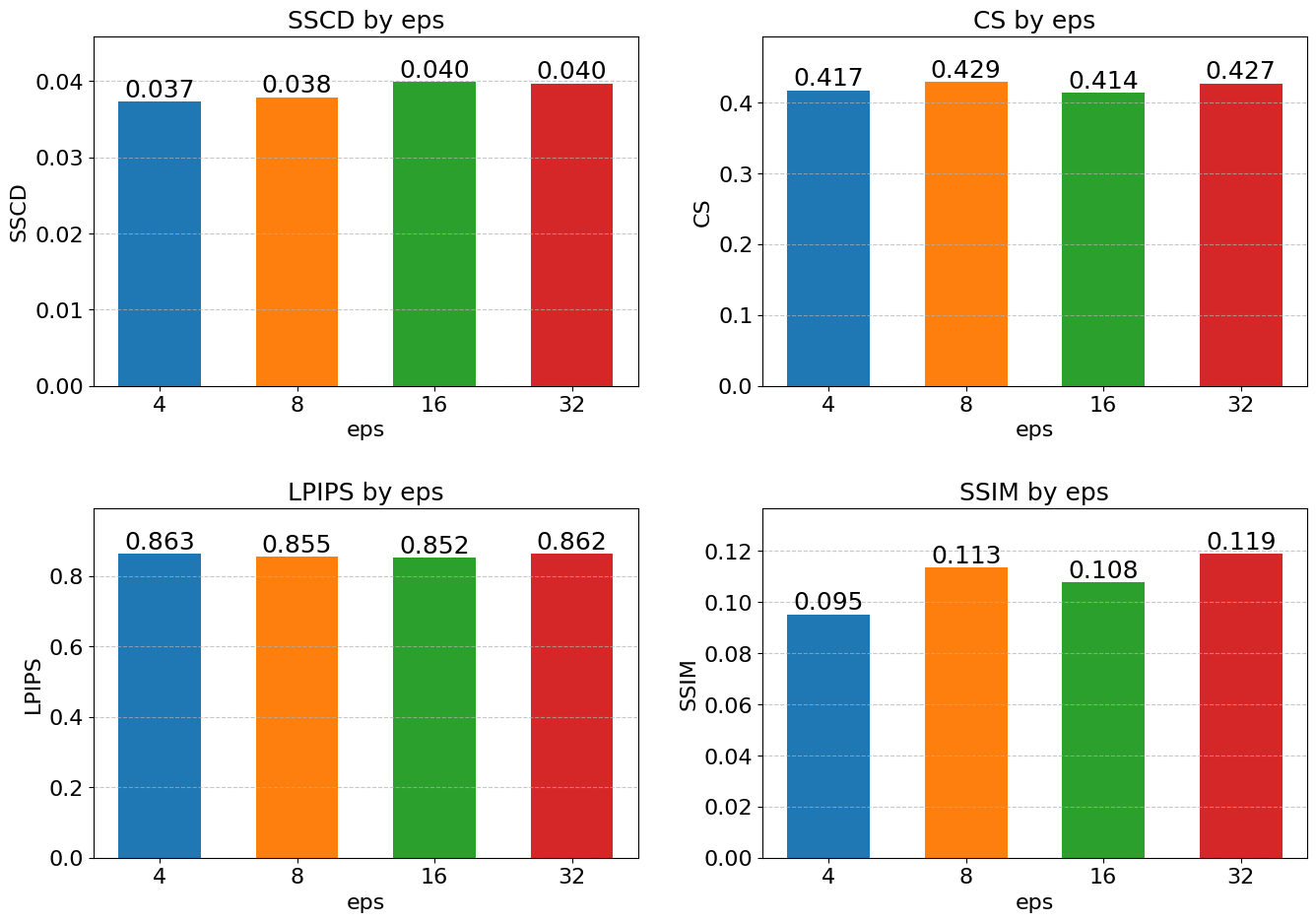}
    \caption{Reconstruction quality results for ViT-L-14 fine-tuned with different values of eps. The influence of eps is minimal within the range of 4 to 32.}
    \label{fig:eps}
\end{figure}

\noindent\textbf{Influence of eps}
To assess the sensitivity of reconstruction quality to the hyperparameter eps, we conduct ablation experiments. Specifically, we fine-tune the model independently using four distinct eps values: 4, 8, 16, and 32.  As visually presented in \cref{fig:eps}, the results clearly indicate that the influence of eps on reconstruction quality is minimal within the tested range of 4 to 32. This suggests that LeakyCLIP is quite robust to the exact setting of eps within this practical range.

\noindent\textbf{More Experiments on Classifiers}
\label{sec:classifiers}
We further evaluate whether AFT is also effective for conventional classifier-based model inversion. 
Following the loss function in \cite{nguyen2023re}, we use KEDMI~\cite{chen2021knowledge} as the inversion method. 
We perform inversion on CelebA~\cite{liu2015deep} and use FFHQ~\cite{karras2019style} as the auxiliary dataset. 
The experiment parameters are controlled via command-line arguments, with PGD selected as the attack type. 
The classifiers are trained using cosine similarity. 
The adversarial perturbations are constrained by the $L_{\infty}$ norm with $\epsilon=2/255$, and PGD is run for 5 steps with a step size of 1.

Since this experiment targets classifier inversion rather than CLIP inversion, we follow the traditional metrics used in classifier-based model inversion. 
Accuracy measures whether the reconstructed image is classified as the target identity by an evaluation classifier, while Accuracy5 measures whether the target identity appears in the top-5 predictions. 
FID evaluates the distributional quality of reconstructed images, where a lower value indicates better visual realism. 
KNN measures the feature-space distance between reconstructed images and real images, where a lower value indicates that reconstructions are closer to real data samples. 
These metrics are standard for classifier inversion because the target information is an identity label. 
However, they are not well-suited for CLIP inversion, where the target condition is a text prompt and the goal is to reconstruct the paired training image. 
Therefore, in our main CLIP experiments, we instead use image-level similarity metrics such as SSIM, LPIPS, CLIP Score, SSCD, and HS.

\begin{table}[h!]
\caption{Performance comparison of classifier-based model inversion with and without AFT.}
\label{tab:methods_comparison}
\centering
\begin{tabular}{l l c c c c}
\toprule
\multirow{2}{*}{\textbf{Model}} 
& \multirow{2}{*}{\textbf{Method}} 
& \multicolumn{4}{c}{\textbf{CelebA/FFHQ}} \\
\cmidrule(r){3-6} 
& 
& Accuracy \(\uparrow\) 
& Accuracy5 \(\uparrow\) 
& FID \(\downarrow\) 
& KNN \(\downarrow\)  \\
\midrule
\multirow{2}{*}{IR152}
& Baseline 
& 68.20 & 90.67 & 51.86 & 1347.90  \\
& AFT (ours) 
& \textbf{72.13} & \textbf{92.67} & \textbf{47.40} & \textbf{1297.75}  \\ 
\midrule
\multirow{2}{*}{FaceNet}
& Baseline  
& 71.07 & 90.00 & 50.95 & 1367.83 \\
& AFT (ours) 
& \textbf{76.27} & \textbf{93.00} & \textbf{46.98} & \textbf{1327.26} \\
\bottomrule
\end{tabular}
\end{table}

For both IR152 and FaceNet, AFT improves all evaluated metrics. 
For IR152, Accuracy increases from 68.20\% to 72.13\%, and Accuracy5 increases from 90.67\% to 92.67\%. 
FID decreases from 51.86 to 47.40, indicating improved image quality, while KNN decreases from 1347.90 to 1297.75, suggesting that the reconstructed images are closer to real samples in feature space. 
Similarly, for FaceNet, Accuracy improves from 71.07\% to 76.27\%, Accuracy5 improves from 90.00\% to 93.00\%, FID decreases from 50.95 to 46.98, and KNN decreases from 1367.83 to 1327.26. 

These results show that AFT is not limited to CLIP inversion. 
It also improves conventional classifier-based inversion by reducing the instability caused by adversarially sensitive features, leading to more effective optimization and higher-quality reconstructions.

\noindent\textbf{Verification of Smoothed Optimization}\;
Gradient norm analysis (\cref{fig:optim_smooth}) demonstrates that AFT yields smaller and less variable gradients during inversion, indicating a smoother optimization landscape. This smoother landscape facilitates more effective optimization, contributing to LeakyCLIP’s enhanced performance.

\noindent\textbf{Distinguishing Extraction from Sampling}\;
To empirically validate that LeakyCLIP performs data extraction rather than mere data generation or sampling, we benchmark it against Stable Diffusion (SD) \cite{rombach2022high}. As shown in \cref{tab:methods_comparison_all}, LeakyCLIP consistently outperforms SD across most of the evaluation metrics.

\begin{figure*}
    \centering
    \includegraphics[width=1.5\columnwidth]{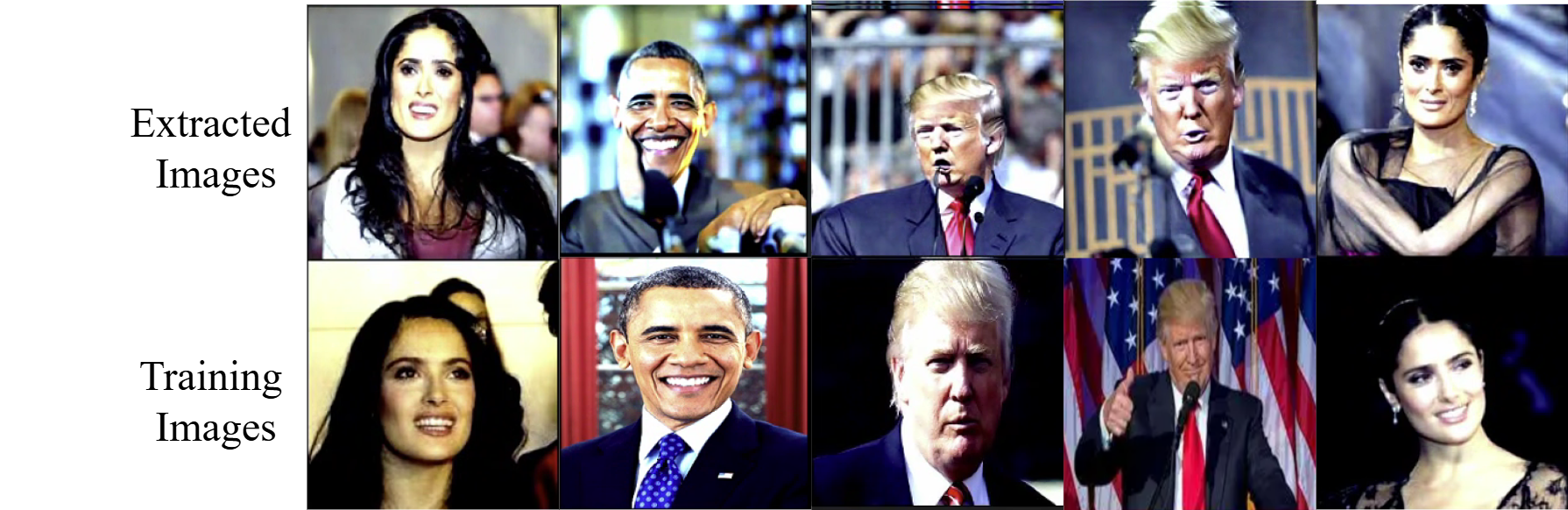}
    
    \caption{Visual comparison of original facial images and their reconstructions.}
    \label{fig:show_face}
\end{figure*}

\noindent \textbf{Extraction of Sensitive Face Data}
To demonstrate LeakyCLIP's capability to extract PII, we reconstruct facial images from LFW \cite{huang2008labeled} and LAION face data (5000 samples total). Results in \cref{tab:facial} and \cref{fig:show_face} show LeakyCLIP successfully reconstructs 5.68\% of faces at HS threshold for ViT-B-16, confirming significant privacy risks for individuals whose data was used for training.
\begin{table}[h!]
\caption{Performance of LeakyCLIP on extracting sensitive face data from the LFW dataset and part of the face data in LAION-2B. }
\centering
\small
\setlength{\tabcolsep}{0.5mm}
\begin{tabular}{llccccc}
\toprule
 & & \multicolumn{5}{c}{\textbf{LFW Dataset+LAION-2B Face Subset }} \\
\cmidrule(lr){3-7}
\textbf{Model} & \textbf{Method} & SSIM $\uparrow$ & LPIPS $\downarrow$ & CS $\uparrow$ & SSCD $\uparrow$ & HS(\%) $\uparrow$ \\
\midrule
\multirow{2}{*}{{ViT-B-16}} 
 & Baseline & 0.042 & 0.833 & 0.320 & 0.018 & 0.000 \\
 & LeakyCLIP & \textbf{0.174} & \textbf{0.753} & \textbf{0.443} & \textbf{0.132} & \textbf{5.680} \\
\midrule
\multirow{2}{*}{{ViT-B-32}} 
 & Baseline & 0.034 & 0.886 & 0.294 & 0.001 & 0.000 \\
 & LeakyCLIP & \textbf{0.163} & \textbf{0.744} & \textbf{0.4583} & \textbf{0.077} & \textbf{2.720} \\
\midrule
\multirow{2}{*}{{ViT-L-14}} 
 & Baseline & 0.036 & 0.987 & 0.2801 & -0.012 & 0.000 \\
 & LeakyCLIP & \textbf{0.105} & \textbf{0.764} & \textbf{0.3591} & \textbf{0.051} & \textbf{1.020} \\
\bottomrule
\end{tabular}

\label{tab:facial}

\end{table}

\noindent\textbf{Membership Inference Protocol}\;
To evaluate whether low-fidelity reconstructions still reveal training membership, we formulate membership inference as a binary classification task. Given a ground-truth image $x$, its caption $t$, and the reconstruction $\hat{x}$ generated by LeakyCLIP, the classifier predicts whether $x$ belongs to CLIP's pre-training corpus. We use LFW as the source of target images and determine membership by searching each image against public LAION-400M CLIP embeddings. Specifically, for each LFW image, we retrieve the top-$k$ nearest LAION candidates with $k=200$, and label the image as a member if at least one candidate satisfies strict near-duplicate criteria based on perceptual hash distance, SSIM, and LPIPS. Thresholds are selected on a separate development set to maintain a low false-positive rate. To reduce identity bias, each member image is paired with a non-member image from the same LFW identity whenever possible, yielding a balanced member/non-member dataset.

For each labeled image, we run the same LeakyCLIP inversion pipeline as in the main experiments, but without diffusion refinement, so that the attack is based on low-fidelity reconstructions rather than visually polished outputs. The inversion model is not trained or fine-tuned on any LFW image used for membership evaluation. For each pair $(x,\hat{x})$, we compute four similarity-based features: SSIM, LPIPS, SSCD, and CLIP image--image cosine similarity. These features define $\phi(x,\hat{x})\in\mathbb{R}^4$. We partition the data at the identity level so that each LFW identity appears in only one split, and evaluate membership prediction with 5-fold cross-validation using logistic regression, random forest, and SVM classifiers.

\noindent\textbf{Privacy Leakage in Low-Fidelity Reconstructions}
To demonstrate the real-world risk of data leakage even in low-fidelity reconstructions, we apply Membership Inference Attack (MIA) \cite{ma2025safety} to low-fidelity reconstructions. LeakyCLIP’s reconstruction metrics, including SSIM, LPIPS, and SSCD, are used as input features for a classifier that predicts whether a given image appears in the original training set. Our experiments focus on CLIP models pretrained on LAION-400M. We draw member samples from the LFW dataset, which consists of human face images and is included within the LAION-400M corpus , while non-member samples comprise faces absent from LAION-400M\cite{hintersdorf2024does}. After selecting only low-fidelity reconstructions (SSCD $<$0.05), we build a training set of 488 samples and a test set of 140 samples. Each split contains equal numbers of images present in the original CLIP training corpus and images absent from it. As shown in \cref{tab:mia_performance}, the MIA demonstrates high effectiveness. This proves that training data membership can be detected even in poor reconstructions, making the privacy risk far more widespread. 

\begin{table}[h!]
\caption{Membership Inference Attack performance across various CLIP models.}
\centering
\small
\setlength{\tabcolsep}{1mm}
\begin{tabular}{lcccccc}
\toprule
\textbf{Method} & \multicolumn{2}{c}{\textbf{\shortstack[c]{Random \\ Forest}}} & \multicolumn{2}{c}{\textbf{\shortstack[c]{Logistic \\ Regression}}} & \multicolumn{2}{c}{\textbf{SVM}} \\
\cmidrule(lr){2-3} \cmidrule(lr){4-5} \cmidrule(lr){6-7}
              & \textbf{Acc} & \textbf{AUC} & \textbf{Acc} & \textbf{AUC} & \textbf{Acc} & \textbf{AUC} \\
\midrule
ViT-B-16 & 0.85 & 0.94 & 0.83 & 0.91 & \textbf{0.89} & \textbf{0.95} \\
ViT-B-32 & 0.86 & 0.93 & 0.84 & 0.92 & \textbf{0.91} & \textbf{0.94} \\
ViT-L-14 & 0.87 & 0.95 & 0.87 & 0.95 & \textbf{0.89} & \textbf{0.97} \\
\bottomrule
\end{tabular} 
\\

\label{tab:mia_performance}

\end{table}
\begin{figure}
    
    \centering
    \includegraphics[width=0.8\linewidth]{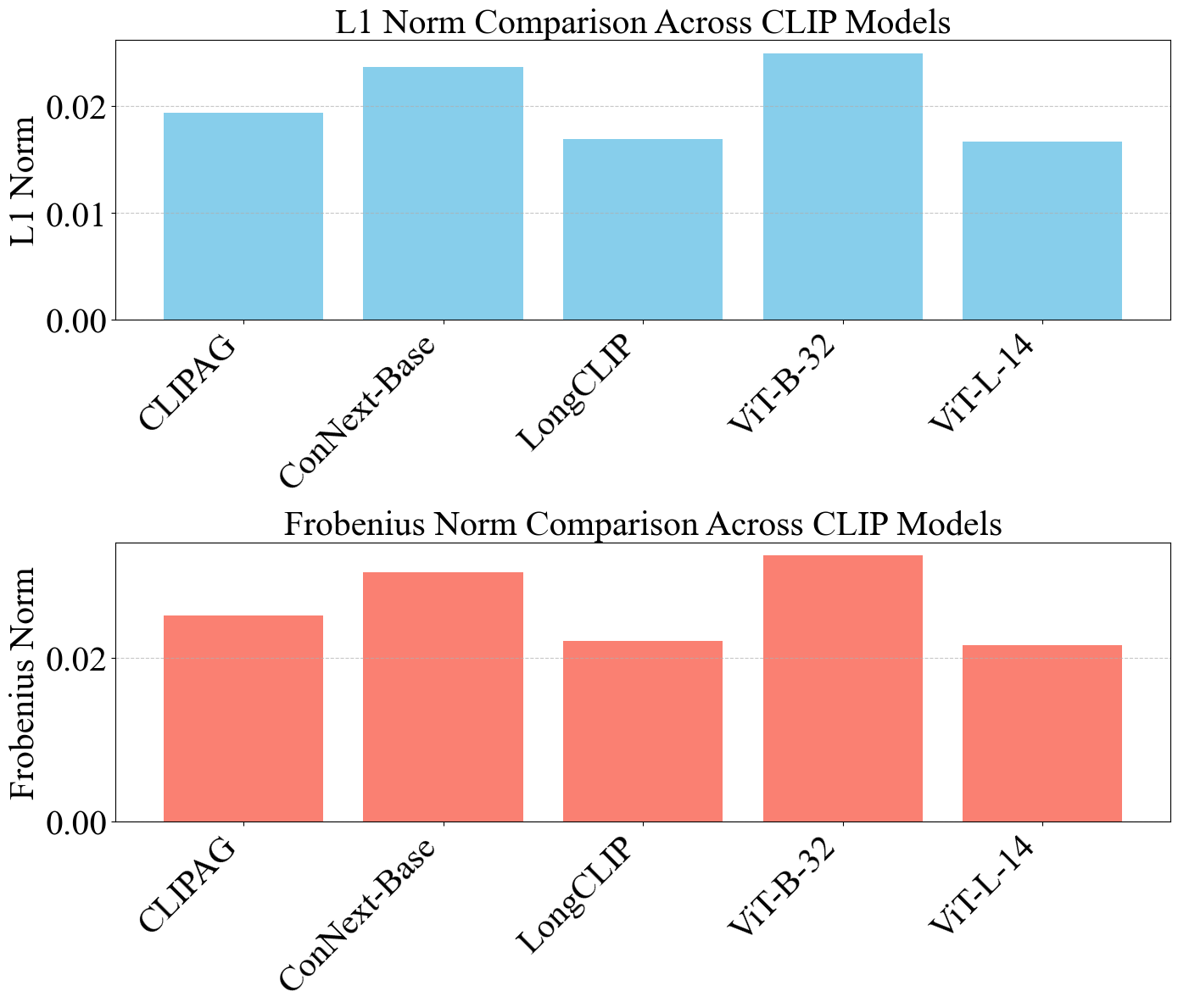}
    
    \caption{Comparison of L1 and Frobenius Norms Across CLIP Models.}
    \label{fig:the_veri}
\end{figure}
\noindent\textbf{Empirical Validation of Linear Relationship}\;
To empirically validate the linear relationship between CLIP text and image embeddings, we learn the mapping matrix $M$ on an auxiliary dataset $D_{\mathrm{aux}}$ using \eqref{M_cal}, and evaluate the reconstructed image embeddings on a disjoint 1495-sample test set from the LAION-HD subset. We conduct this evaluation across LongCLIP~\cite{zhang2025long}, CLIPAG~\cite{ganz2024clipag}, ConvNext-Base, ViT-B-32, and ViT-B-16. As shown in \cref{fig:selection,fig:the_veri}, using only 2,000 image--text pairs to learn $M$ already yields an average cosine similarity of approximately $0.8$ between reconstructed and original image embeddings. The reconstruction errors are also small, with an average L1 error of approximately $0.02$ and a Frobenius-norm error of around $55$ relative to the full $1495 \times \mathrm{dim}$ target matrix. These results support the existence of a stable linear relationship between CLIP text and image embeddings and show that the learned mapping generalizes well from a small auxiliary dataset.

\noindent\textbf{Empirical Validation of Theorem~\ref{thm:AFT}}\;
Theorem~\ref{thm:AFT} states that the reconstruction error induced by noisy inputs is controlled by the Lipschitz condition number of the image encoder:
\begin{equation}
    W_2^2(p,\hat{p}) \leq 4d\sigma^2 \left(\frac{M}{m}\right)^2 ,
\end{equation}
where $M$ is the upper Lipschitz constant and $m$ is the lower Lipschitz constant. This bound suggests that inversion quality depends not only on encoder smoothness, but also on whether the ratio $M/m$ remains small. Intuitively, $M$ controls the largest possible feature change caused by a small image perturbation, whereas $m$ measures whether visually different inputs remain distinguishable in the embedding space. A desirable encoder should therefore reduce $M$ without substantially collapsing $m$.

\begin{figure*}[t]
    \centering
    \includegraphics[width=\textwidth]{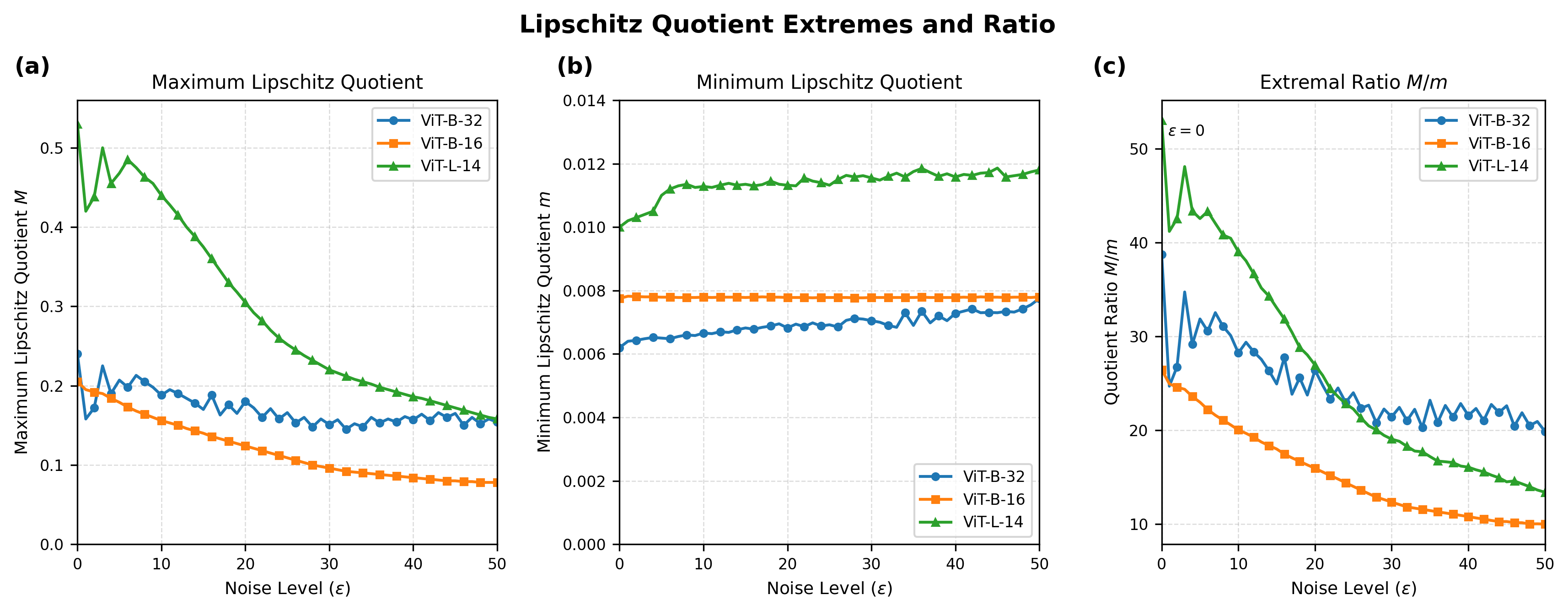}
    \caption{
    Empirical validation of Theorem~\ref{thm:AFT}. 
    Left: estimated maximum Lipschitz quotient $M$ under different perturbation levels. 
    Middle: estimated minimum Lipschitz quotient $m$. 
    Right: the resulting extremal ratio $M/m$. 
    The results show that the upper Lipschitz quotient varies much more strongly than the lower quotient, indicating that the main source of inversion instability comes from large local feature sensitivity. Since the theoretical reconstruction bound scales as $(M/m)^2$, reducing $M$ while preserving $m$ yields a smoother optimization landscape and tighter reconstruction bound.
    }
    \label{fig:lipschitz_ratio}
\end{figure*}

To empirically validate this mechanism, we estimate the local maximum and minimum Lipschitz quotients on 500,000 images sampled from LAION-2B under different perturbation levels. As shown in \cref{fig:lipschitz_ratio}, the maximum quotient $M$ varies substantially across models and noise levels, while the minimum quotient $m$ remains comparatively stable. This indicates that the dominant source of instability in CLIP inversion comes from the upper Lipschitz envelope rather than from a loss of feature distinguishability. In particular, a large $M$ means that small pixel-level perturbations can induce disproportionately large changes in the image embedding, producing noisy and highly variable gradients during inversion. Since the inversion objective directly depends on the encoder output $f_{\mathrm{FT}}(x)$, reducing this worst-case local sensitivity makes the optimization landscape smoother.

The ratio plot in \cref{fig:lipschitz_ratio} further supports the theoretical prediction. Because the reconstruction bound scales quadratically with $M/m$, even a moderate reduction in $M$ can substantially tighten the bound when $m$ is approximately preserved. This explains why adversarial fine-tuning improves CLIP inversion: AFT suppresses the worst-case local sensitivity of the encoder, thereby reducing $M$ and smoothing the optimization landscape, while largely maintaining the lower Lipschitz behavior needed to preserve discriminative visual information. Consequently, AFT makes gradient-based inversion more stable and reduces noise-induced feature fluctuations, consistent with the improved reconstruction quality observed in our experiments.

\begin{figure*}[!t]
    \centering

    \begin{minipage}{0.48\textwidth}
        \centering
        \includegraphics[height=0.5\textheight]{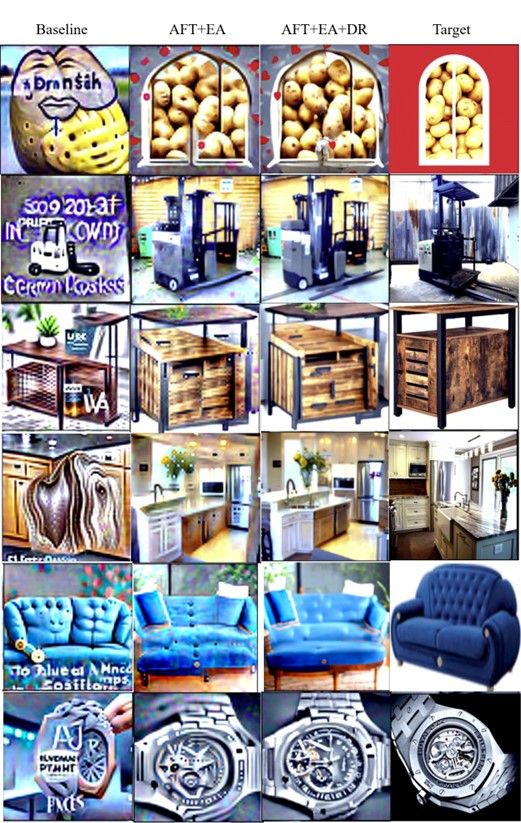}
    \end{minipage}
    \hfill
    \begin{minipage}{0.48\textwidth}
        \centering
        \includegraphics[height=0.5\textheight]{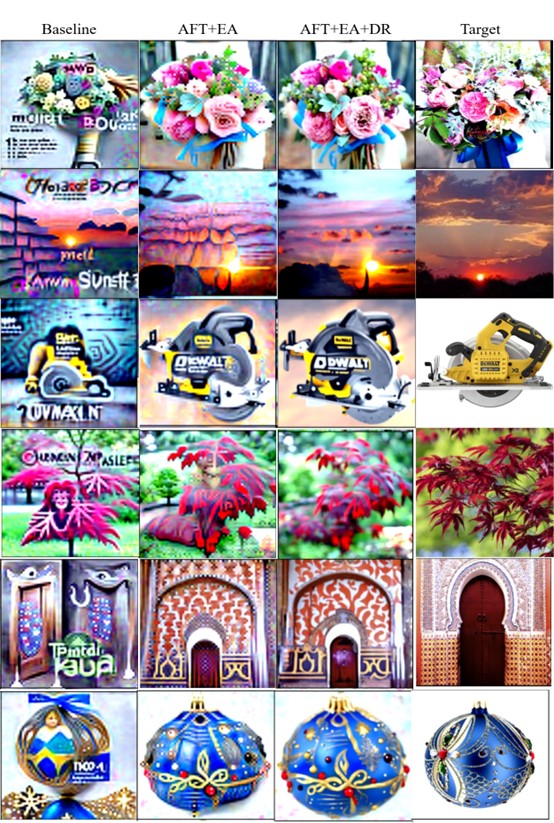}
    \end{minipage}

    \caption{
    Additional qualitative comparison. From left to right in each example:
    AFT, AFT+EA, AFT+EA+DR, and the target image.
    }

    \label{fig:visual_examples}
\end{figure*}

\noindent\textbf{Verifying No Data Leakage from Diffusion Refinement}\;
To ensure reconstructed details originate from CLIP embeddings rather than SD memorization, we compared direct SD sampling using ground-truth captions against our full pipeline. As shown in \cref{tab:methods_comparison_all} (``SD Sampling''), SD alone achieves lower reconstruction scores. Qualitative comparisons in \cref{fig:visual_examples} show images before and after DR are highly similar except for low-level features, confirming SD refines CLIP-provided structure rather than leaking its own memorized content.





\noindent\textbf{Additional Visual Results}\;
We provide additional qualitative comparisons in \cref{fig:visual_examples}. 
Each example shows the reconstruction produced by AFT, AFT+EA, and the full 
AFT+EA+DR pipeline, together with the target image. The results show that EA improves 
the semantic alignment of the AFT reconstruction, while DR further enhances local 
fidelity and visual realism. These examples provide qualitative evidence that the 
LeakyCLIP components progressively improve reconstruction quality. Interestingly, in complex scenes (e.g., snow-capped mountains or bowls of citrus fruits), DR not only enhances fine texture but also corrects subtle geometric distortions introduced by AFT+EA, suggesting that diffusion refinement contributes to high-frequency detail. Across all six generated strips, the same trends are observed, highlighting the generality of LeakyCLIP across diverse visual domains (natural landscapes, still lifes, objects, and vehicles).

\noindent\textbf{Black-Box Setting Evaluation} We extend LeakyCLIP to the black-box setting where target model parameters are inaccessible. To the best of our knowledge, there are no baselines in this setting. We use ViT-B-16 (pretrained on LAION-400M) as a surrogate model, fine-tuning it on auxilary dataset using embeddings from the target model (ConvNext-Base-W trained on LAION-Aesthetic). We then perform inversion on deduplicated LAION-Aesthetic samples.

As shown in Table~\ref{tab:blackbox}, black-box extraction achieves lower but non-trivial performance. HS drops from 4.20\% (white-box on ViT-B-16) to 0.82\% (black-box targeting ConvNext-Base-W), representing expected degradation since the surrogate only approximates the target's embedding space. However, the non-zero HS rate confirms that even approximate embedding inversion leaks member information, emphasizing privacy concerns in realistic deployment scenarios. This result aligns with prior work on model inversion and shows that embedding-space approximations suffice to partially reconstruct memorized content.

\begin{table}[h!]
\caption{Black-box extraction results.}
\centering
\resizebox{0.5\textwidth}{!}{%
\begin{tabular}{lcccccc}
\toprule

\textbf{Setting} & \textbf{Model} & \textbf{SSIM} $\uparrow$ & \textbf{LPIPS} $\downarrow$ & \textbf{CS} $\uparrow$ & \textbf{SSCD} $\uparrow$ & \textbf{HS(\%)} $\uparrow$ \\
\midrule
White & ViT-B-16 & 0.151 & 0.819 & 0.462 & 0.065 & 4.20 \\
White & ConvNext & 0.138 & 0.825 & 0.451 & 0.058 & 3.68 \\
Black & ConvNext & 0.092 & 0.871 & 0.384 & 0.029 & 0.82 \\
\bottomrule
\end{tabular}
}
\label{tab:blackbox}
\end{table}

\begin{figure}[h]
    \centering
    \includegraphics[width=0.95\columnwidth]{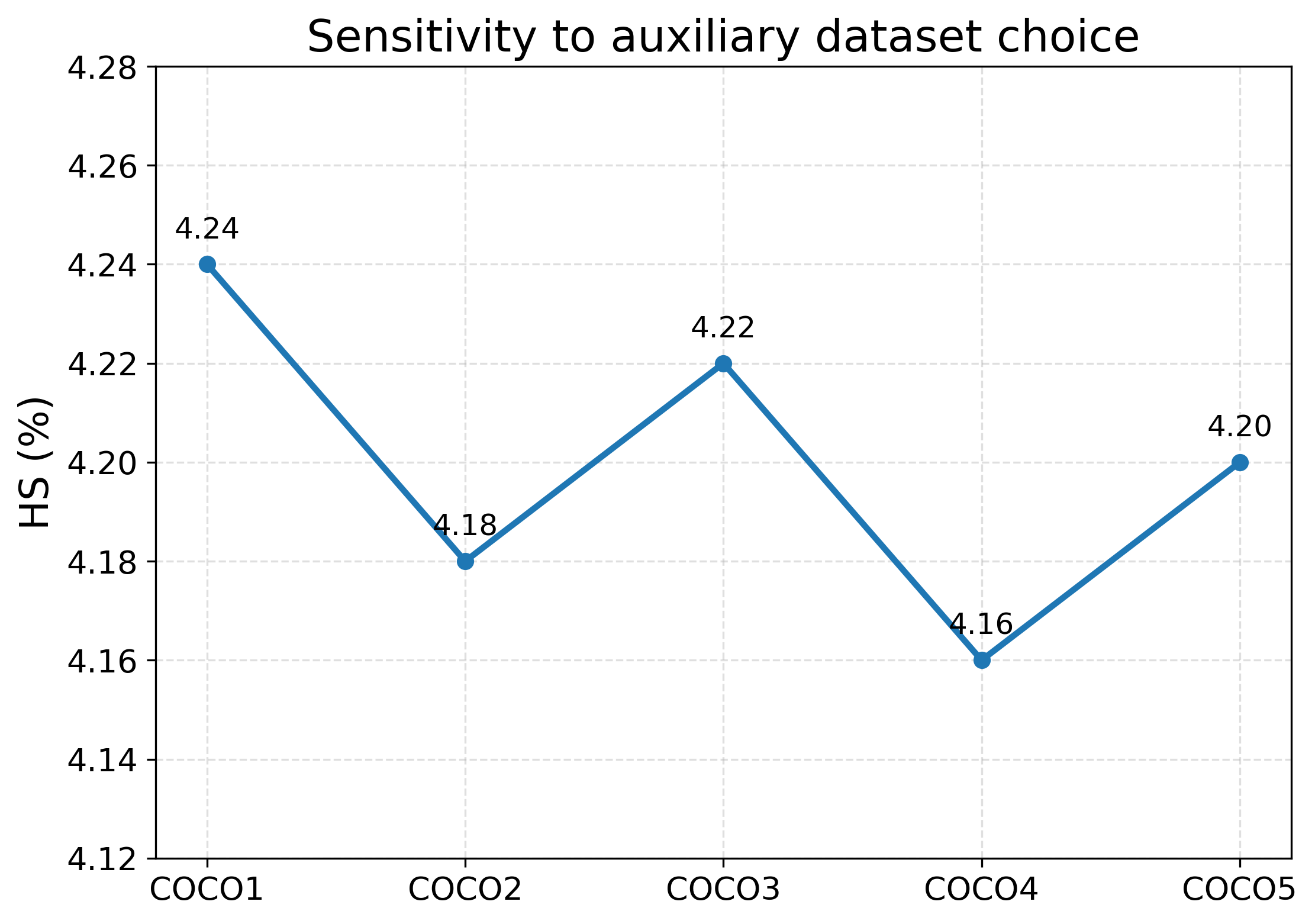}
    \caption{Sensitivity to auxiliary dataset choice. HS rates remain stable when using five different non-overlapping COCO subsets as the auxiliary dataset for embedding alignment.}
    \label{tab:coco_aux_hs_cols}
\end{figure}
\begin{figure}[h]
    \centering
    \includegraphics[width=0.95\columnwidth]{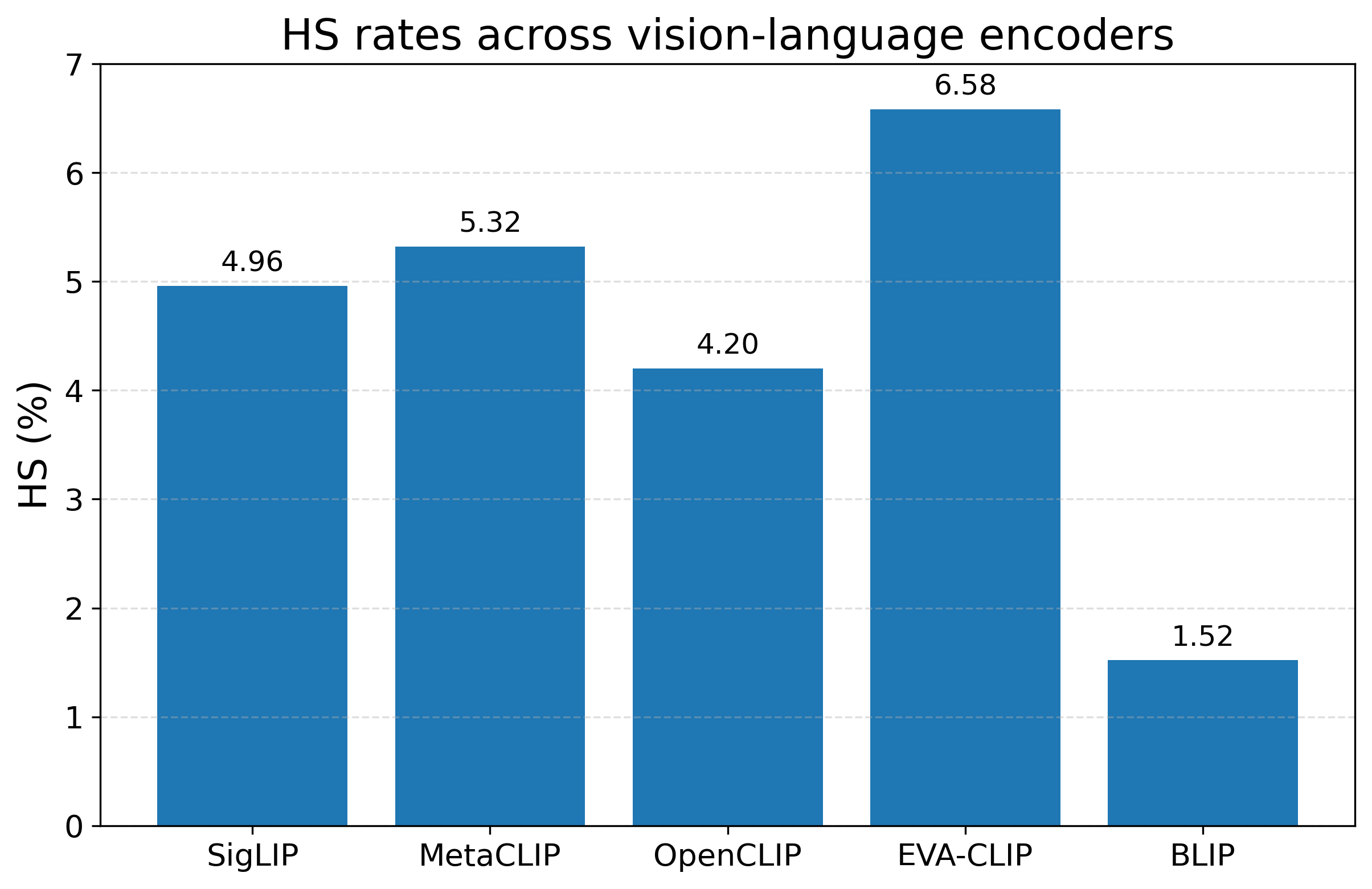}
    \caption{HS rates across additional vision-language encoders on the LAION-2B subset. LeakyCLIP achieves non-zero HS rates across all evaluated encoders.}
    \label{tab:hs_rate}
\end{figure}

\begin{figure}[!t]
    \centering
    \includegraphics[width=0.95\columnwidth]{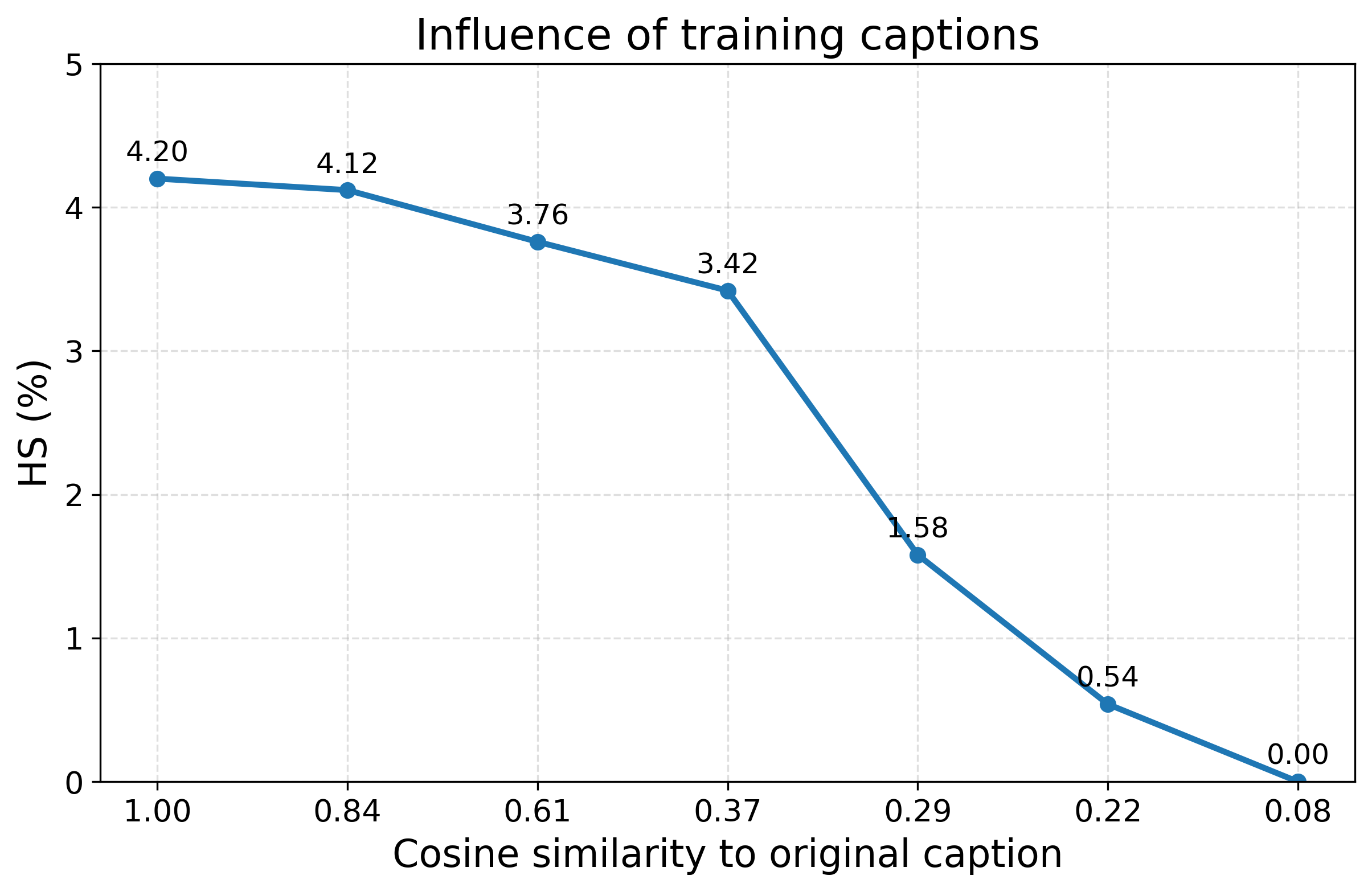}
    \caption{Effect of caption rewriting on HS rates. Cosine similarity is computed between the rewritten and original CLIP text embeddings.}
    \label{tab:caption_shift_hs}
\end{figure}

\noindent\textbf{Sensitivity to Auxiliary Dataset Choice}\;
We evaluate whether LeakyCLIP is sensitive to the auxiliary dataset used for embedding alignment. As shown in Fig.~\ref{tab:coco_aux_hs_cols}, replacing the original auxiliary set with five COCO subsets yields minimal variation in HS (4.16–4.24\%), a max difference of 0.08 pp. This indicates that the linear mapping $M$ estimated during EA is robust to specific dataset choices, as long as sufficient diversity and coverage of image-text pairs are maintained. The stability underscores that leakage arises from embedding structure rather than memorization of auxiliary images. Notably, variance is slightly higher for more complex scenes (e.g., architecture or multi-object still lifes), suggesting subtle dataset biases can still influence reconstruction fidelity.

\noindent\textbf{Evaluation on More Vision-Language Encoders}\;
We further evaluate LeakyCLIP on additional vision-language encoders, including SigLIP, MetaCLIP, OpenCLIP, EVA-CLIP, and BLIP. As shown in \cref{tab:hs_rate}, LeakyCLIP achieves non-zero HS rates across all tested encoders, suggesting that training-data leakage is not specific to a single CLIP implementation. BLIP shows a lower HS rate, likely because its mixed pretraining objectives make direct CLIP-style inversion less effective.

\noindent\textbf{Influence of Training Captions}\;
We also study the effect of text-domain shift by rewriting the original training captions with GPT-4o and measuring the cosine similarity between the rewritten and original CLIP text embeddings. As shown in \cref{tab:caption_shift_hs}, the HS rate decreases as the rewritten caption moves farther from the original prompt. Nevertheless, LeakyCLIP can still extract non-trivial high-fidelity samples when the rewritten prompt has only $0.37$ cosine similarity to the original caption. This suggests that the leakage risk persists under moderate caption variation, but becomes substantially weaker under severe text-domain shift.

\section{Conclusion}

In conclusion, this paper introduces LeakyCLIP, a novel and effective attack for extracting training data from CLIP models. LeakyCLIP addresses key challenges in CLIP inversion through adversarial fine-tuning, embedding alignment, and Diffusion-based refinement. Empirical results demonstrate LeakyCLIP's superior reconstruction quality. Furthermore, we proved that the privacy risk is pervasive, showing that training data membership can even be successfully inferred from low-fidelity reconstructions. Our findings confirm that data leakage is a practical and widespread vulnerability in CLIP, highlighting the urgent need for more robust defenses in the training and deployment of CLIP models.


%
%
\bibliographystyle{splncs04}
\bibliography{main}

\onecolumn
\newpage
\appendices

\section{Theoretical Analysis and Understanding}
\label{sec:thm}
In this section, we analyze the theoretical feasibility of recovering training images from text embeddings, assuming the image encoder is a bijection. We also discuss how detailed textual descriptions amplify the risk of training data leakage. 

\subsection{Notations and Setting}

\label{sec:setting}

\hspace{1em}Let $\mathcal{X}$ denote the distribution space of the training images, with 
$p(x)$ representing the corresponding probability distribution. The dataset $\mathcal{D}=\{x_i\}$ consists of i.i.d. samples drawn from 
$p(x)$. Given a label $Y$, we can partition $\mathcal{D}$ into several disjoint subsets $\mathcal{D}_k$, each with a distribution $p_k(x)$. The original distribution can be seen as a mixture of these component distributions: $p(x) = \sum \pi_k p_k(x)$, where $\pi_k = P(Y = k)$. We refer to labels that satisfy these properties as \textbf{informative labels}.

For CLIP models, let $\theta$ denote the model parameters trained on $\mathcal{D}$, and let $p_{\theta}(x)$ represent the distribution of images generated by CLIP inversion. When we restrict the generated images to a specific label $k$, the distribution of these images is denoted as $p_{\theta_k}(x)$. The image encoder $f_I: \mathcal{X} \xrightarrow{}\mathcal{Z}$ maps images from the original space $\mathcal{X}$ to a latent space $\mathcal{Z}$. The distribution of image representations in the latent space $\mathcal{Z}$ is denoted by $q(z)$ and $q_{k}(z)$ for the respective subsets. We use $\mu$(and $\mu_k$) to denote expectations, and $\Sigma$(and $\Sigma_k$) for the covariance matrices of $q(z)$(and $q_k(z)$). We use the following  point-wise memorization metric defined in \cite{chen2024towards} to quantify the data extraction performance.

\begin{definition}
\label{def:mem-metric}
{\normalfont\bfseries (Point-wise Memorization)}
The point-wise memorization of a generative model $f_{\theta}$ with respect to its training samples $\mathcal{D} = \{x_i\}_{i = 1}^{| \mathcal{D} |}$ is defined as: 
\begin{align}
\label{eq:point-wise}
    \mathcal{M}_{point}(\mathcal{D};p_\theta) = \frac{1}{|\mathcal{D}|} \sum_{i = 1}^{|\mathcal{D}|} \int p_\theta(x) \log{\frac{p_\theta(x)}{q(x;x_i,\epsilon)}} dx,
\end{align}
where $p_\theta$ denotes the distribution of data generated by the model, and $q(x;x_i;\epsilon) \sim \mathcal{N}(x_i,\epsilon I)$ is a normal distribution centered at training sample $x_i$ with small variance $\epsilon$ .
\end{definition}



\subsection{Theoretical Analysis}

\label{sec:theorical-results}
\begin{theorem}
\label{thm:memory}
    Assume the  total variation distance between $p_k(x)$ and $p_{\theta_k}(x)$ is 0, namely $TV(p_k(x),p_{\theta_k}(x)) = 0$ and the image encoder $f_I(x):\mathcal{X} \rightarrow \mathcal{Z}$ is a continuous differentiable bijection and normalizes the representations to the unit sphere(i.e., $\mathcal{Z} \subset \mathcal{S}^{m-1}$), for each dataset $\mathcal{D}_k$ with $||\mu||_2 < ||\mu_k||_2$, with probability 1, we have: 
    \begin{equation}
        \lim_{|\mathcal{D}_k| \rightarrow \infty} 
        \lim_{\epsilon \rightarrow 0} \frac{\mathcal{M}_{point}(\mathcal{D}_k ; q_k)}{\mathcal{M}_{point}(\mathcal{D}_k ; q)} < 1.
    \end{equation}
    When $||\mu_k||_2 \geq \frac{||\mu||_2+\sqrt{||\mu||^2_2+ \eta(2+\eta)}}{2+\eta} $ holds for some  $\eta > 0$, then with probability 1, we have:
    \begin{equation}
        \label{eq:strong}
        \lim_{|\mathcal{D}_k| \rightarrow \infty} 
        \lim_{\epsilon \rightarrow 0} \frac{\mathcal{M}_{point}(\mathcal{D}_k ; q_{k)}}{\mathcal{M}_{point}(\mathcal{D}_k ; q)} \leq \frac{2}{2+\eta} < 1.
    \end{equation}
\end{theorem}
Under these assumptions, Theorem~\ref{thm:memory}
characterizes how properties of the (idealized) encoder and
caption length jointly influence the invertibility of the mapping.

We emphasize that this result should be interpreted as
\emph{heuristic intuition} for our method rather than a precise
description of CLIP’s behavior. Specifically, the condition $|| \mu ||_2 < || \mu_k ||_2$ is not difficult to meet because:
\begin{align}
    tr(\Sigma_k) = \mathbb{E}[||z||_2^2]- ||\mathbb{E}[z]||_2^2 = 1-||\mu_k||_2^2.
\end{align}
Thus, this condition can be interpreted as constraining the variance of $q_k(z)$ in the latent space, which can be achieved with a careful selection of the informative label.
Theorem \ref{thm:memory} is consistent with prior works \cite{somepalli2023understanding,chen2024towards}. They show that unconditional models don't replicate data, while text-conditioning increases memorization. And previous work \cite{gu2023memorization} demonstrates that random-label conditioning enhances memorization.

Adding multiple informative labels to the text description is equivalent to sequentially partitioning the dataset multiple times. Under the assumptions in Theorem~\ref{thm:memory}, we can apply it iteratively, yielding Corollary~\ref{cor:1}:
\begin{corollary}
\label{cor:1}
Given $n$ informative labels, let $p_{k_{1...n}}(x)$ denote the distribution of images satisfying these labels, and $q_{k_{1...n}}(z)$ denote the distribution of $f_I(x)$ where $x \sim p_{k_{1...n}}$. If each partition satisfies the assumptions in  Theorem~\ref{thm:memory}, then with probability 1:
\begin{equation}
        \label{eq:n-descriptions}
        \lim_{|\mathcal{D}_{k_{1...n}}| \rightarrow \infty} \lim_{\epsilon \rightarrow 0} \frac{\mathcal{M}_{point}(\mathcal{D}_{k_{1...n}} ; q_{k_{1...n}})}{\mathcal{M}_{point}(\mathcal{D}_{k_{1...n}} ; q)} \leq (\frac{2}{2+\eta})^n.
    \end{equation}
\end{corollary}

Corollary \ref{cor:1} is empirically supported by previous literature \cite{carlini2023extracting,kiyomaru-etal-2024-comprehensive-analysis}. They observe log-linear growth of extractable sequences with token count, which is named the "discoverability phenomenon".

When the number of labels $n$ is sufficiently large, the data distribution that satisfies these conditions in the latent space $\mathcal{Z}$ will converge to a single point. This can occur by encoding all the information about images. In this case, we can omit the $|\mathcal{D}_{k_{1...n}}| \rightarrow \infty$ condition in Corollary~\ref{cor:1}, resulting in:
\begin{align}
    \label{image_embedding}
    \lim_{n \rightarrow \infty}\lim_{\epsilon \rightarrow 0} \frac{\mathcal{M}_{point}(\mathcal{D}_{k_{1...n}} ; q_{k_{1...n}})}{\mathcal{M}_{point}(\mathcal{D}_{k_{1...n}} ; q)}\rightarrow 0.
\end{align}

The corollaries of Theorem~\ref{thm:memory} \emph{suggest} that,
within this simplified model, increasing the number of tokens
in the caption tends to improve the success probability of
inversion. We view these statements as qualitative guidance
for designing our attack (e.g., encouraging richer textual
descriptions), rather than as guarantees that hold exactly for
real CLIP encoders.


\paragraph{Proof of Theorem~\ref{thm:memory}}
\label{pf:memory}
By assumptions of Theorem~\ref{thm:memory}, $TV(p_k(x),p_{\theta_k}(x)) = 0$ and $f_I$ is a continuous differentiable bijection $q(z) = \sum_k \pi_k q_k(z)$ almost everywhere. Now, we compute the point-wise memorization of $q_k$ on dataset $\mathcal{D}_k$ here:
\begin{equation}
\begin{split}
    \mathcal{M}_{point}(\mathcal{D}_k ; q_k) &= \frac{1}{|\mathcal{D}_k|} \sum_{i = 1}^{|\mathcal{D}_k|} \int q_k(z) \log{\frac{q_k(z)}{q(z;z_i,\epsilon)}} dz  \\
    &= -H(q_k) + \frac{m}{2}\log{2\pi\epsilon} +\frac{1}{|\mathcal{D}_k|} \sum_{i = 1}^{|\mathcal{D}_k|} \int q_k(z) \frac{(z-z_i)^T(z-z_i)}{2\epsilon} dz \\
    &= -H(q_k) + \frac{m}{2}\log{2\pi\epsilon} + \frac{1}{2 \epsilon |\mathcal{D}_k|}\sum_{i = 1}^{|\mathcal{D}_k|}(tr(\Sigma_k) + (\mu_k - z_i)^T(\mu_k - z_i)) \\
    &\rightarrow o(\frac{1}{\epsilon}) + \frac{1}{\epsilon} tr(\Sigma_k) \ \text{almost surely as} \ |\mathcal{D}_k| \rightarrow \infty \text{ by the strong law of large number}
\end{split}
\end{equation}

where for each sample $z_i$:
\begin{equation}
\begin{split}
&\int q_k(z) \frac{(z-z_i)^T(z-z_i)}{2\epsilon} dz \\
&= \int  \frac{q_k(z)}{2\epsilon} \left((\mu_k-z_i)^T(\mu_k-z_i) + 2(\mu_k-z_i)^T(z-\mu_k) + (z-\mu_k)^T(z-\mu_k)\right)dz \\
&=\frac{1}{2\epsilon}(tr(\Sigma_k) + (\mu_k-z_i)^T(\mu_k-z_i))
\end{split}
\end{equation}

Hence as $\epsilon \rightarrow0$, $|\mathcal{D}_k| \rightarrow \infty$:
\begin{equation}
\begin{split}
    \mathcal{M}_{point}(\mathcal{D}_k ; q_k) &= 
     o(\frac{1}{\epsilon})+ \frac{1}{2 \epsilon |\mathcal{D}_k|}\sum_{i = 1}^{|\mathcal{D}_k|}(tr(\Sigma_k) + (\mu_k - z_i)^T(\mu_k - z_i)) \\
    &\rightarrow o(\frac{1}{\epsilon}) + \frac{1}{\epsilon} tr(\Sigma_k) \ \text{almost surely by the strong law of large number}
\end{split}
\end{equation}

Similarly, for $\mathcal{M}_{point}(\mathcal{D}_k ; q)$:
\begin{equation}
\begin{split}
    \mathcal{M}_{point}(\mathcal{D}_k ; q) &= o(\frac{1}{\epsilon}) + \frac{1}{2 \epsilon |\mathcal{D}_k|}\sum_{i = 1}^{|\mathcal{D}_k|}(tr(\Sigma) + (\mu - z_i)^T(\mu - z_i)) \\
    &=o(\frac{1}{\epsilon}) + \frac{1}{2 \epsilon |\mathcal{D}_k|}\sum_{i = 1}^{|\mathcal{D}_k|}(tr(\Sigma) + ||\mu - \mu_k||_2^2 + 2(\mu_k - z_i)^T(\mu - \mu_k) + ||\mu_k - z_i)||^2_2)\\
    &\rightarrow o(\frac{1}{\epsilon})+\frac{1}{2\epsilon} (tr(\Sigma) + tr(\Sigma_k) + ||\mu - \mu_k||^2_2)
\end{split}
\end{equation}

Finally, we have
\begin{equation}
\begin{split}
    \lim_{|\mathcal{D}|_k \rightarrow \infty} \lim_{\epsilon \rightarrow 0} \frac{\mathcal{M}_{point}(\mathcal{D}_k ; q_k)}{\mathcal{M}_{point}(\mathcal{D}_k ; q)} &= \frac{2tr(\Sigma_k)}{tr(\Sigma_k) + tr(\Sigma) + ||\mu - \mu_k||^2_2} 
\end{split}
\end{equation}

\begin{equation}
\begin{split}
        \Sigma = \underset{z \sim q(z)}{\mathrm{Cov}}(z) &= \mathbb{E}_q [zz^T] - \mathbb{E}_q[z] \mathbb{E}_q[z]^T\\
        &=\sum_k \pi_k \mathbb{E}_{q_k} [zz^T] - \mu \mu^T \\
        &=\sum_k \pi_k (\mu_k \mu_k^T + \Sigma_k) - \mu \mu^T \\
        &= \mathbb{E}_{\pi}[\Sigma_j] + \mathbb{E}_{\pi} [ (\mu - \mu_j)(\mu - \mu_j)^T]
\end{split}
\end{equation}

Here we use $\mathbb{E}_{\pi}[tr(\Sigma_j)]$ and $\mathbb{E}_{\pi}[||\mu-\mu_j||_2^2]$ to denote $\sum_j \pi_j tr(\Sigma_j)$ and $\sum_j \pi_j ||\mu-\mu_j||_2^2$.

\indent Clearly, by Jensen's inequality we have
\begin{equation}
    \mathbb{E}_\pi \left[\lim_{|\mathcal{D}_k| \rightarrow \infty} \lim_{\epsilon \rightarrow 0} \frac{\mathcal{M}_{point}(\mathcal{D}_k ; q_k)}{\mathcal{M}_{point}(\mathcal{D}_k ; q)}\right] \leq 1
\end{equation}

The equality holds only when $\mu_k$ and $tr(\Sigma_k)$ are the same.

We note that: 
\begin{equation}
\begin{split}
    &tr(\Sigma) + ||\mu - \mu_k||^2_2 - (1+\eta)tr(\Sigma_k) \\
    &=1-||\mu||_2^2  + ||\mu - \mu_k||^2_2 - (1+\eta)(1-||\mu_k||_2^2) \\
    &= (2+\eta)||\mu_k||_2^2 - 2\mu^T\mu_k -\eta \\
    &\geq (2+\eta)||\mu_k||_2^2 - 2||\mu||_2 ||\mu_k||_2 -\eta
\end{split}
\end{equation}
\indent Hence $||\mu_k||_2 \geq \frac{||\mu||_2+\sqrt{||\mu||^2_2+ \eta(2+\eta)}}{2+\eta}$ implies  $tr(\Sigma) + ||\mu - \mu_k||^2_2 \geq (1+\eta)tr(\Sigma_k)$

That leads to:
\begin{equation}
    \lim_{|\mathcal{D}_k| \rightarrow \infty} \lim_{\epsilon \rightarrow 0} \frac{\mathcal{M}_{point}(\mathcal{D}_k ; q_k)}{\mathcal{M}_{point}(\mathcal{D}_k ; q)} \leq \frac{2}{2+\eta}
\end{equation}

As for Corollary~\ref{cor:1}, we only need to note:
\begin{align}
    \frac{\mathcal{M}_{point}(\mathcal{D}_{k_{1...n}} ; q_{k_{1...n}})}{\mathcal{M}_{point}(\mathcal{D}_{k_{1...n}} ; q)} = \prod_{j = 1}^n\frac{\mathcal{M}_{point}(\mathcal{D}_{k_{1...n}} ; q_{k_{1...j}})}{\mathcal{M}_{point}(\mathcal{D}_{k_{1...n}} ; q_{k_{1...j-1}})}
\end{align}

And by Theorem~\ref{thm:memory} for each $j \in \{1,2,...,n\}$:
\begin{align}
    \frac{\mathcal{M}_{point}(\mathcal{D}_{k_{1...n}} ; q_{k_{1...j}})}{\mathcal{M}_{point}(\mathcal{D}_{k_{1...n}} ; q_{k_{1...j-1}})} \leq \frac{2}{2+\eta}
\end{align}

In addition, consider the sequence $\{a_n\}$ such that $a_1 = \sqrt{\frac{\eta}{2+\eta}}$ and $a_{k+1} = \frac{a_k+\sqrt{a_k^2+ \eta(2+\eta)}}{2+\eta}$. By the conditions in Corollary~\ref{cor:1}, we have $a_j \leq \mu_{k_{1...j}} \leq 1$ for any positive integer $j$. It's easy to verify that $\{a_n\}$ is monotonically increasing and converges to 1. So we have $\displaystyle \lim_{n \rightarrow \infty} ||\mu_{k_{1...n}||_2} = 1$ and then the covariance matrix $\Sigma_{k_{1...n}} \rightarrow 0$, which means that when $n$ is sufficiently large, the representations of images satisfying all of the $n$ informative labels will distribute in a small region with high probability.

\subsection{Empirical Validation for bijection}
\label{ssec:empirical_bi_injection}

In Theorem~\ref{thm:memory}, the CLIP image encoder $f_I$ is assumed to be a differentiable bijection. While differentiability follows from the neural network structure, we empirically verify the bijectivity assumption by showing that $f_I$ satisfies the bi-Lipschitz property.



\begin{definition}
\label{def:bi-Lipschitz}
{\normalfont\bfseries (Bi-Lipschitz Property)}
A function \( f : \mathcal{X} \to \mathcal{Z} \), where 
\( \mathcal{X} \subseteq \mathbb{R}^m, \mathcal{Z} \subseteq \mathbb{R}^n \), 
is called \textbf{bi-Lipschitz} if there exist constants 
\( 0 < L_1 \le L_2 < \infty \) such that for all 
\( x_1, x_2 \in \mathcal{X} \),
\[
    L_1\, \|x_1 - x_2\|_2
    \le \|f(x_1) - f(x_2)\|_2
    \le L_2\, \|x_1 - x_2\|_2.
\]
\end{definition}
The bi-Lipschitz condition implies that 
$f$ is injective and admits a Lipschitz continuous inverse on its image, hence establishing a bijection between $\mathcal{X}$ and $f(\mathcal{X})$. To empirically verify the property, we calculate the \textbf{Lipschitz quotient}, $L(x_1, x_2)$, which provides a local estimate for the Lipschitz constants:
\begin{equation}
L(x_1, x_2) = \frac{\|f_I(x_1) - f_I(x_2)\|_2}{\|x_1 - x_2\|_2}
\label{eq:Lipschitz-quotient}
\end{equation}
We investigate this property on a large-scale dataset of \textbf{10 million natural images} randomly sampled from the LAION-400M dataset~\cite{schuhmann2022laion}. For each model (ViT-B/32, ViT-B/16, and ViT-L/14), we take an image $x$ and create a perturbed version $x' = x + \delta$, where $\delta$ is random Gaussian noise of a specified magnitude (Noise Level).

\begin{figure}[ht!]
    
    \centering
    \includegraphics[width=\textwidth]{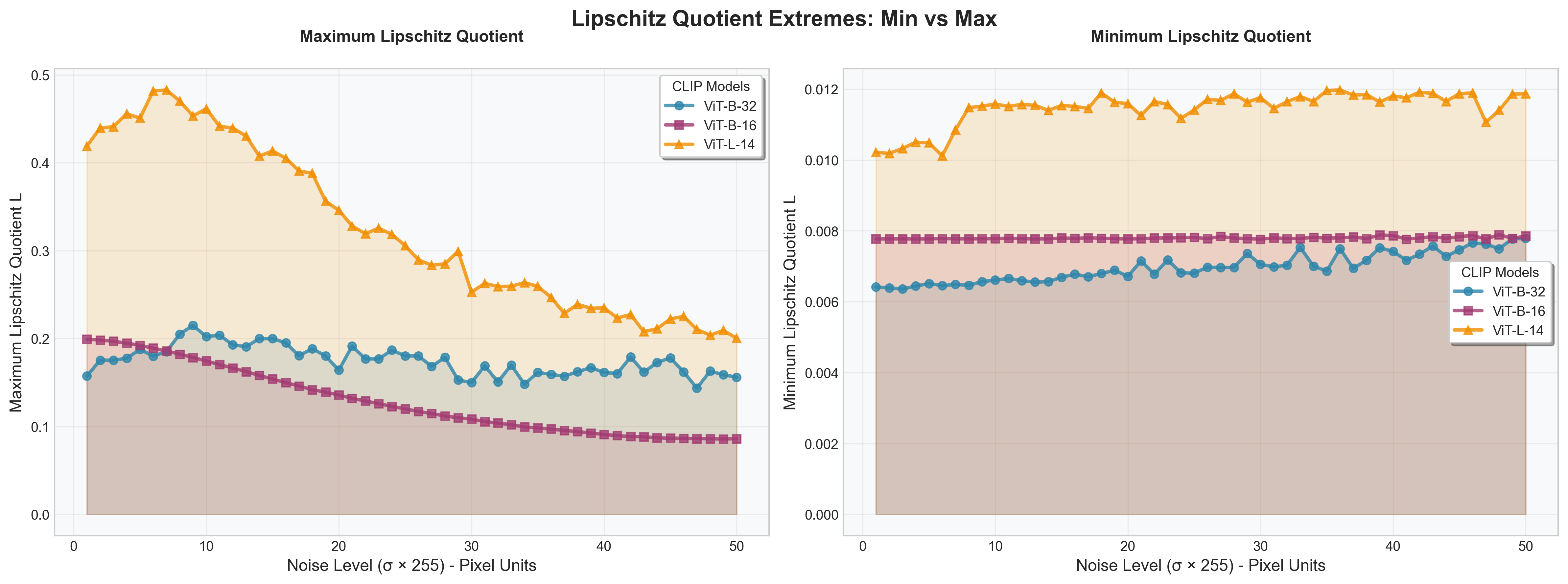}
    
    \caption{\textbf{Lipschitz Quotient Extremes Across CLIP Models}. The $\sigma$ shown in the X-axis is multiplied by 255}
    \label{fig:Lipschitz_quotient}
\end{figure}

The results are presented in Figure~\ref{fig:Lipschitz_quotient}. The left panel displays the maximal observed Lipschitz quotient, providing an empirical bound for the upper constant $L_2$. The right panel shows the minimal Lipschitz quotient, which estimates the lower constant $L_1$. The critical observation is that across all tested models and noise levels, the \textbf{minimal Lipschitz quotient remains strictly greater than zero}.  Consequently, the bi-Lipschitz condition holds, providing an empirical foundation for our assumptions.

\subsection{Empirical Validations for Theorem~\ref{thm:memory} and Corollary~\ref{cor:1}}
\paragraph{Experiment Settings}
To validate Theorem ~\ref{thm:memory} and Corollary~\ref{cor:1}, we design the following experiments. Firstly, we randomly sample 1,000 images from the Laion-2B dataset \cite{schuhmann2022laion}. For each image, we use Gemini-1.5-pro to generate a description of the image containing tokens ranging from 1 to 25. Then, we use IL to evaluate the quality of images reconstructed from the text embeddings corresponding to these descriptions.

By Corollary~\ref{cor:1}, the inversion loss would decrease exponentially as the number of tokens increases. We fit a linear regression model of the form:
\begin{align}
\log{\mathcal{L}} = \alpha + \beta \cdot T + \epsilon,
\end{align}
where $\mathcal{L}$ represents the inversion loss, $T$ denotes the number of tokens, $\alpha$ and $\beta$ denote the regression coefficients, and $\epsilon$ is the error term. According to Corollary~\ref{cor:1}, we expect a negative slope $\beta$, reflecting the theoretical bound $\log(\frac{2}{2+\eta})$.

\begin{figure}
    
    \centering
    \includegraphics[width=0.55\linewidth]{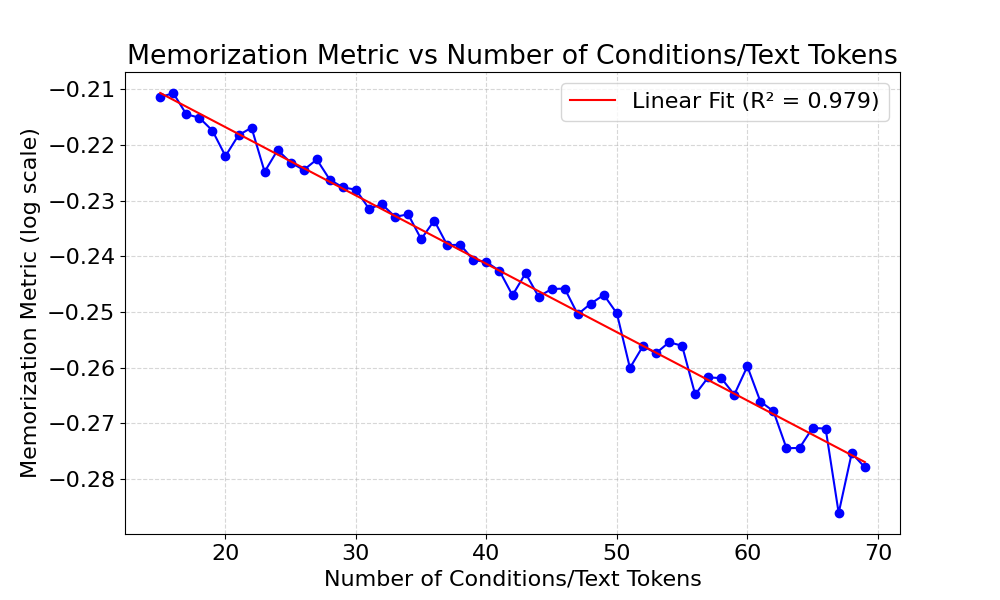}
    
    \caption{Empirical validation of
Theorem \ref{thm:memory} and Corollary \ref{cor:1}. The plot depicts the relationship between the number of conditions (representing the details of text prompts) and the Memorization Metric (log scale).}
    \label{fig:th1_empirical}
\end{figure}

\begin{table}
    \caption{OLS Regression Results. The results confirm the theoretical predictions of Theorem \ref{thm:memory} and Corollary \ref{cor:1}.}
    \centering
    
    \begin{tabular}{lc}
        \toprule
        \textbf{Variable} & \textbf{Coefficient (Std. Error)} \\
        \midrule
        $\alpha$ & -0.1173 (0.011)*** \\
        $\beta$ & -0.0228 (0.001)*** \\
        \midrule
        \multicolumn{2}{l}{\textbf{Model Statistics}} \\
        \midrule
        R-squared & 0.469 \\
        F-statistic & 992.0 \\
        Prob (F-statistic) & 1.47e-156 \\
        Durbin-Watson & 1.762 \\
        \bottomrule
    \end{tabular}
    \begin{flushleft}
        \textit{Note:} *** indicates $p < 0.001$.
    \end{flushleft}
    
    \label{tab:regression_results}
\end{table}

\begin{figure}
    
    \centering
    \includegraphics[width=0.55\linewidth]{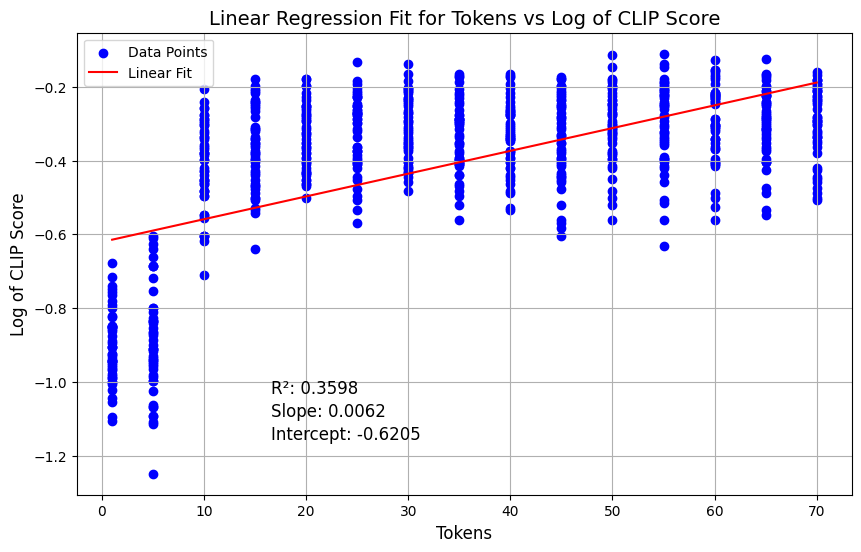}
    
    \caption{The relationship between the number of tokens and the logarithm of inversion loss. This result empirically verifies Theorem \ref{thm:memory}
 and Corollary \ref{cor:1}.}
    \label{fig:th1_empirical_2}
\end{figure}

\paragraph{Results}

The regression results presented in Table~\ref{tab:regression_results} and \cref{fig:th1_empirical} reveal a negative coefficient of \(-0.0228\) (p \(<\) 0.001) between $T$ and $\log{\mathcal{L}}$.
These results indicate that the fitted linear model is robust and consistent with the theoretical predictions, providing strong empirical support for the hypothesized relationships.

\subsection{Proof of Theorem~\ref{thm:AFT}}
\begin{proof}
Let
\[
\tilde{x}=x+\xi,
\ \text{where}  \ \mathbb{E}{\| \xi \|^2} \leq d \sigma^2 \]
and define the reconstruction
\[
\hat{x}
=
\arg\min_{z}
\|f_I(z)-f_I(\tilde{x})\|^2.
\]
By the optimality of $\hat{x}$,
\[ \|f_I(\hat{x})-f_I(\tilde{x})\|
\le \|f_I(x)-f_I(\tilde{x})\|. \]
Since $f_I$ is $L$-Lipschitz,
\[ \|f_I(x)-f_I(\tilde{x})\| =
\|f_I(x)-f_I(x+\xi)\| \le
M\|\xi\|. \]
Hence,
\[ \|f_I(\hat{x})-f_I(\tilde{x})\|
\le M\|\xi\|. \]
Applying the triangle inequality yields
\[ \|f_I(\hat{x})-f_I(x)\|
\le \|f_I(\hat{x})-f_I(\tilde{x})\|
+ \|f_I(\tilde{x})-f_I(x)\|
\le 2M\|\xi\|. \]
Using the lower bi-Lipschitz condition, we obtain
\[ \|\hat{x}-x\|
\le \frac{1}{m} \|f_I(\hat{x})-f_I(x)\|
\leq \frac{2M}{m} \| \xi \|\]
Let $\gamma=\mathrm{Law}(x,\hat{x})$ denote the joint distribution induced by $(x,\hat{x})$. Since the marginals of $\gamma$ are $p$ and $\hat{p}$,  we have
$\gamma\in\Pi(p,\hat{p})$where $\Pi(p,\hat{p})$ denotes the set of all couplings between $p$ and $\hat{p}$.By the definition of the $2$-Wasserstein distance,
\[ W_2^2(p,\hat{p})
= \inf_{\mu\in\Pi(p,\hat{p})}
\mathbb{E}_{(x,y)\sim\mu}\|x-y\|^2.\]
Since $\gamma$ is a feasible coupling,
\[W_2^2(p,\hat{p})
\le \mathbb{E}_{(x,\hat{x})\sim\gamma}
\|x-\hat{x}\|^2. \]
Combining this with the previous inequality, we complete the proof 
\[
W_2^2(p,\hat{p}) \le
\frac{4M^2}{m^2} \mathbb{E}\|\xi\|^2
\leq \frac{4d\sigma^2M^2}{m^2}
\]

\end{proof}

\subsection{Proof of Theorem~\ref{th:text2image}}

We analyze whether the mapping from text embeddings \( \mathbf{u}_T \) to image embeddings \( \mathbf{u}_I \) in CLIP can be represented as a single linear transformation. Starting from the graph-based formulation of CLIP, we derive a \emph{coupled linear relation} between the stacked text and image eigen-embeddings. This relation is linear in \( \mathbf{U}_T \) and \( \mathbf{U}_I \), but, as we will see, it depends on all nodes jointly and therefore cannot, in general, be reduced to one matrix acting independently on each individual text embedding to produce its paired image embedding.

\paragraph*{Eigenvalue Decomposition}

We perform eigenvalue decomposition on the (normalized) Laplacian \( \mathbf{L} \):
\begin{align}
\mathbf{L} = \mathbf{U} \Lambda \mathbf{U}^\top ,
\end{align}
where:
\begin{itemize}
    \item \( \mathbf{U} \) is the matrix of eigenvectors, partitioned according to text and image nodes as
    \begin{align}
    \mathbf{U} =
    \begin{pmatrix}
    \mathbf{U}_T \\
    \mathbf{U}_I
    \end{pmatrix},
    \end{align}
    with \( \mathbf{U}_T \in \mathbb{R}^{m \times (m+n)} \) corresponding to text nodes and \( \mathbf{U}_I \in \mathbb{R}^{n \times (m+n)} \) corresponding to image nodes.
    \item \( \Lambda \) is the diagonal matrix of eigenvalues.
\end{itemize}

\paragraph*{Constructing Embeddings}

We select the top \( d \) eigenvectors corresponding to the smallest non-zero eigenvalues to construct embeddings of dimension \( d \). Denote by
\(\Lambda_d \in \mathbb{R}^{d \times d}\) the diagonal matrix of the selected eigenvalues, and by
\(\mathbf{U}_T \in \mathbb{R}^{m \times d}\), \(\mathbf{U}_I \in \mathbb{R}^{n \times d}\) the corresponding truncated eigenvector matrices (for notational simplicity we reuse the same symbols). The embeddings are defined as
\begin{align}
\mathbf{u}_T &= \mathbf{U}_T \Lambda_d^{1/2} \in \mathbb{R}^{m \times d}, \\
\mathbf{u}_I &= \mathbf{U}_I \Lambda_d^{1/2} \in \mathbb{R}^{n \times d}.
\end{align}

From the eigenvalue equation \( \mathbf{L} \mathbf{U} = \mathbf{U} \Lambda \) and the block structure of the normalized Laplacian on the bipartite graph, we obtain
\begin{align}
\begin{pmatrix}
\mathbf{I}_m & -\mathbf{D}_T^{-1/2} \mathbf{W} \mathbf{D}_I^{-1/2} \\
-\mathbf{D}_I^{-1/2} \mathbf{W}^\top \mathbf{D}_T^{-1/2} & \mathbf{I}_n
\end{pmatrix}
\begin{pmatrix}
\mathbf{U}_T \\
\mathbf{U}_I
\end{pmatrix}
=
\begin{pmatrix}
\mathbf{U}_T \Lambda \\
\mathbf{U}_I \Lambda
\end{pmatrix}.
\end{align}

This leads to two coupled equations. For the image nodes, we have
\begin{align}
-\mathbf{D}_I^{-1/2} \mathbf{W}^\top \mathbf{D}_T^{-1/2} \mathbf{U}_T
+ \mathbf{I}_n \mathbf{U}_I
= \mathbf{U}_I \Lambda .
\end{align}
Rearranging terms and restricting to the selected \( d \) dimensions yields
\begin{align}
\mathbf{U}_I (\mathbf{I}_d - \Lambda_d)
= \mathbf{D}_I^{-1/2} \mathbf{W}^\top \mathbf{D}_T^{-1/2} \mathbf{U}_T ,
\end{align}
which is exactly the relation stated in \cref{eq:linear_relation}. This equation shows that, after graph-based reweighting through
\(\mathbf{D}_T\), \(\mathbf{D}_I\), and \(\mathbf{W}\), the image eigen-embeddings lie in the linear span of the text eigen-embeddings.

However, note that the operator \( \mathbf{D}_I^{-1/2} \mathbf{W}^\top \mathbf{D}_T^{-1/2} \) acts on the \emph{entire} matrix \( \mathbf{U}_T \) and depends on the global graph structure and spectrum. Thus, while the relation above is linear in \( \mathbf{U}_T \) and \( \mathbf{U}_I \), it does not in general correspond to a single matrix that can be applied independently to each individual text embedding to yield its paired image embedding. This is why, in the main text, we introduce a data-driven surrogate matrix \( M \) and learn it via least squares as an approximation to this coupled operator on a given dataset. This completes the proof of Theorem~\ref{th:text2image}.

\end{document}